\documentclass[12pt,a4paper,oneside,openany]{book}
\usepackage[margin=27mm,headheight=28pt]{geometry}
\usepackage[T1]{fontenc}
\usepackage{lmodern}
\usepackage{amsmath,amssymb,mathtools}
\usepackage{graphicx}
\usepackage{booktabs,longtable,array,multirow,tabularx}
\usepackage{enumitem}
\usepackage{caption,subcaption}
\usepackage{xcolor}
\usepackage[hidelinks]{hyperref}
\usepackage{microtype}
\usepackage{fancyhdr}
\graphicspath{{thesis_translation/figures/}}
\setlist{nosep,leftmargin=2.2em}
\makeatletter
\def\@makechapterhead#1{%
  \vspace*{0\p@}%
  {\parindent\z@\raggedright\normalfont
    \ifnum\c@secnumdepth>\m@ne
      \Large\bfseries \@chapapp\space \thechapter\par\nobreak
      \vskip 4\p@
    \fi
    \interlinepenalty\@M
    \LARGE\bfseries #1\par\nobreak
    \vskip 12\p@}}
\def\@makeschapterhead#1{%
  \vspace*{0\p@}%
  {\parindent\z@\raggedright\normalfont
    \interlinepenalty\@M
    \LARGE\bfseries #1\par\nobreak
    \vskip 12\p@}}
\makeatother
\newcommand{\thesisTitle}{Research on Deep Learning-Based Semantic Segmentation Algorithms for Subcortical Brain Structures}

\begin{document}
\pagenumbering{gobble}
\begin{titlepage}
\centering
\vspace*{1.7cm}
\vspace{2.2cm}
{\Large MASTER'S THESIS\par}
\vspace{1.8cm}
{\bfseries\LARGE \thesisTitle\par}
\vfill
\begin{tabularx}{0.92\textwidth}{@{}l >{\raggedright\arraybackslash}X@{}}
Institution: & State Key Laboratory of Medical Molecular Biology, Institute of Basic Medical Sciences, Peking Union Medical College, Tsinghua University\\
Candidate: & Binbin Yang\\
Supervisor: & Weiwei Zhang\\
Advisory committee member: & Jian Liu\\
Discipline: & Biomedical Engineering\\
Research area: & Pattern Recognition and Medical Image Processing\\
Completion date: & 11 May 2020\\
\end{tabularx}
\end{titlepage}

\frontmatter

\chapter*{Abstract}
\addcontentsline{toc}{chapter}{Abstract}
Segmentation of subcortical brain structures is fundamental to computer-aided diagnosis and treatment in neurology and related clinical fields. Segmenting and analyzing brain structures in magnetic resonance (MR) images can support the early diagnosis and treatment of conditions such as autism spectrum disorder, stroke, and brain tumors. To improve the accuracy of subcortical structure segmentation, this thesis develops two related deep learning approaches for brain MR images: DenseMedic and the alternate connected neural network (ACNN).

The first part of the work develops DenseMedic. First, the OreoDown method accelerates receptive-field growth by introducing larger convolutional strides at earlier layers and restores network depth by interleaving size-preserving convolutional layers, thereby translating faster receptive-field growth into an effective increase in receptive field. Second, DenseMedic instantiates the OreoDown framework using the construction principle of DenseNet and obtains multiscale contextual information through densely connected feature-extraction operations. Finally, hybrid dilated convolutions are introduced at individual layers to enlarge the receptive field further and refine otherwise coarse feature perception. The second part develops ACNN. First, alternate connections between layers are proposed to instantiate OreoDown as a single-path ACNN. Second, the single path is divided centrally to form a multipath ACNN without changing the stated parameter count, providing a unified architecture for single- and multimodal segmentation. Finally, prior information is introduced into the multipath ACNN for single-modal images; images with slightly different intensity distributions are supplied to separate paths, further improving performance on single-modal tasks.

Experiments were conducted on the public IBSR and MRBrainS18 datasets for subcortical brain segmentation. Performance was evaluated using the Dice similarity coefficient (DSC), intersection over union (IoU), 95th-percentile Hausdorff surface distance (HSD95), and average surface distance (ASD). Ablation experiments on both datasets evaluated the proposed components. Comparisons with contemporary algorithms assessed DenseMedic on single-modal T1-weighted MR images and ACNN on single-modal T1-weighted, dual-modal T1-weighted plus FLAIR, and dual-modal T1-weighted plus T1-IR tasks. The experimental results showed greater regional overlap and closer boundary agreement between the predicted and reference structures, supporting the accuracy and robustness of both methods for the evaluated subcortical structures. In clinical applications, accurate segmentation of these structures may facilitate measurement of diagnostic indicators and accelerate computer-assisted treatment planning.

\noindent\textbf{Keywords:} subcortical brain structures; semantic segmentation; OreoDown; DenseMedic; alternate connected neural network

\cleardoublepage
\tableofcontents
\cleardoublepage
\listoffigures
\cleardoublepage
\listoftables

\mainmatter
\chapter{Introduction}

\section{Background, Purpose, and Significance}

\begin{figure}[tbp]
\centering
\includegraphics[width=0.62\linewidth]{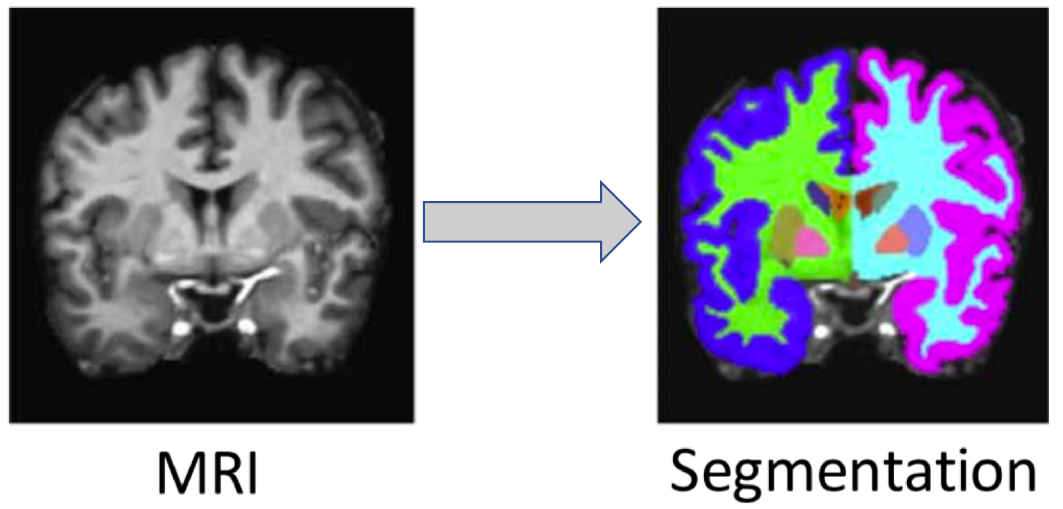}
\caption{Schematic illustration of subcortical brain structures.}
\label{fig:subcortical-structures}
\end{figure}

The human brain comprises the cerebrum, cerebellum, and brainstem. As the highest center of the nervous system, it has highly complex structures and functions. Figure~\ref{fig:subcortical-structures} illustrates the subcortical brain structures. The cerebral cortex contains the principal concentration of neurons; approximately 14 billion neurons form a complex system of connections that regulates bodily functions and provides the basis for human cognition. Cerebral white matter contains millions of neural fibers composed of ensheathed axons. These fibers connect neurons in different cortical and subcortical regions and transmit action potentials between them. The cerebral cortex has a complex regional organization, and individual subcortical structures are involved in the precise control of different bodily functions. Clear and accurate delineation of these structures therefore enables quantitative studies of brain function and can advance neuroscience as well as the diagnosis and treatment of neurological disease. Improvements in medical technology and imaging equipment have substantially increased the quality of brain imaging in recent years. Researchers in neuroscience and clinical medicine can now delineate tissue regions in brain images and perform noninvasive quantitative studies of subcortical structures and their functions, thereby supporting the diagnosis and treatment of brain disorders.

\begin{figure}[tbp]
\centering
\includegraphics[width=0.68\linewidth]{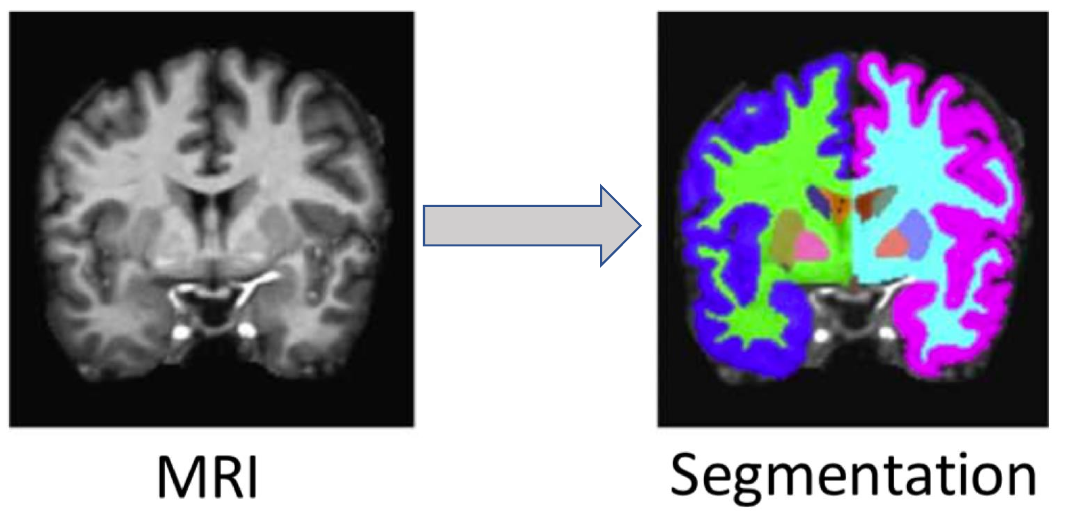}
\caption{Schematic illustration of subcortical brain structure segmentation.}
\label{fig:segmentation-concept}
\end{figure}

Magnetic resonance imaging (MRI) has become a preferred modality for brain imaging because it involves no ionizing radiation, is noninvasive, permits multiplanar reconstruction, and provides excellent soft-tissue contrast. If every voxel in a brain MR image can be classified appropriately, the cerebral cortex, cerebral white matter, and other subcortical structures can be delineated, thereby achieving subcortical segmentation. As illustrated by the coronal section in Figure~\ref{fig:segmentation-concept}, a class label is assigned to each voxel such that voxels belonging to the same brain structure share the same label\cite{ref1}. The MR image is consequently partitioned into multiple nonoverlapping regions according to the underlying structures. Accurate delineation of these regions facilitates measurement of key diagnostic indicators and supports research on brain function and computer-assisted treatment.

In cognitive neuroscience, accurate subcortical segmentation enables more precise measurement of the location, shape, cortical thickness, and volume of different brain tissues. Reliable structural measurements are important across many areas of brain research. Accurate brain-image segmentation supports quantitative studies of schizophrenia\cite{ref2}, Parkinson's disease\cite{ref3}, autism\cite{ref4}, and multiple sclerosis\cite{ref5}, and is also important for evaluating structural abnormalities\cite{ref6}. For example, changes in the shape and volume of the caudate nucleus and putamen are strongly associated with autism spectrum disorder and may assist in assessing restrictive and repetitive behaviors\cite{ref7}. In clinical neurology, rapid and accurate segmentation of brain tissues and structures may also support the diagnosis of disorders such as stroke and brain tumors and facilitate timely treatment planning\cite{ref8,ref9}. Accurate subcortical segmentation can therefore improve understanding of complex neurological disorders and contribute to advances in clinical care.

The accurate segmentation algorithms developed in this thesis may also be relevant to the diagnosis and treatment of several brain disorders. In autism spectrum disorder, morphological and volumetric changes in the caudate nucleus and putamen are closely related to diagnosis; precise segmentation can improve measurement of these structures and thereby support the assessment of restrictive and repetitive behaviors. In radiotherapy for brain tumors, clinicians manually delineate target volumes and organs at risk on patient images so that radiation dose is concentrated within the target while surrounding healthy tissues and organs are spared. Voxel-by-voxel delineation is time-consuming. Accurate segmentation of normal brain regions could help optimize dose distributions and their spatial placement, facilitate automatic delineation of radiotherapy targets, reduce treatment costs, and preserve valuable treatment time. Improvements in subcortical segmentation performance are therefore potentially important for clinical applications.

\section{Review of Related Research}

\subsection{Development and Current Status of Brain Structure Segmentation}

Brain structures were traditionally delineated manually, voxel by voxel, by specialist clinicians. In small datasets, annotations from multiple experts were often fused to form the reference standard. Manual delineation, however, requires substantial knowledge of neuroanatomy, is labor-intensive and time-consuming, and can delay treatment. It is also subjective, and annotations may differ across observers\cite{ref20}. These limitations restrict the scalability of manual voxel-level annotation. Researchers have consequently proposed many automatic methods for subcortical segmentation, including atlas-based methods\cite{ref21,ref22}, statistical-model-based methods\cite{ref23,ref24}, deformable models\cite{ref25}, and machine-learning methods\cite{ref26,ref27}.

\begin{figure}[tbp]
\centering
\begin{subfigure}{0.93\linewidth}
\centering\includegraphics[width=\linewidth]{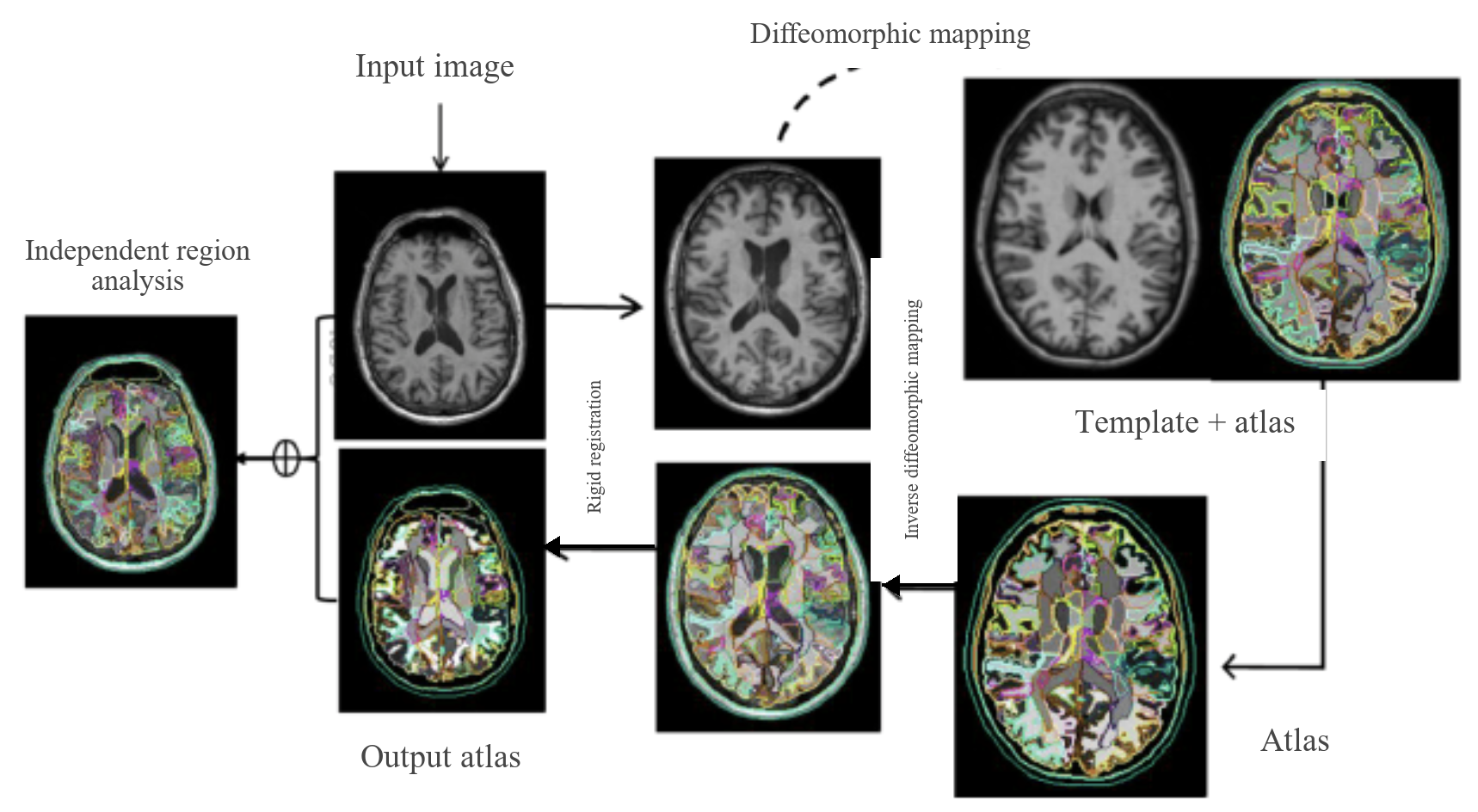}
\caption{Atlas-based segmentation.}
\end{subfigure}
\par\medskip
\begin{subfigure}{0.86\linewidth}
\centering\includegraphics[width=\linewidth]{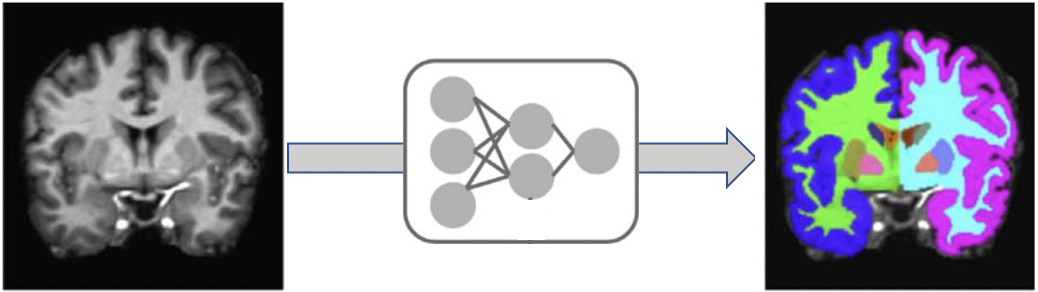}
\caption{End-to-end deep-learning segmentation.}
\end{subfigure}
\caption{Two segmentation paradigms.}
\label{fig:segmentation-paradigms}
\end{figure}

As shown in Figure~\ref{fig:segmentation-paradigms}a, atlas-based segmentation registers an image to a previously segmented template and then maps the template labels back into the original image space using the inverse transformation. Linear and nonlinear registration are commonly described as rigid and nonrigid registration, respectively. Atlas-based methods often produce satisfactory results, but image-to-template registration can involve complex and lengthy procedures. High-resolution images or images with complex structural deformation may require hours or even days to segment. Common brain templates may also be inaccurate for unusual images, including scans from young subjects with incompletely developed brain structures or subjects with structural abnormalities. Severe misalignment or deformation may prevent successful registration altogether.

Statistical-model-based methods use training data to learn parametric representations of structural properties such as shape and texture. Because the parameters are iteratively updated by searching around an initial configuration, accurate initialization is required for convergence to a local optimum. If the number of training images is small relative to the number of learned parameters, overfitting and biased results may occur. Noise in the training data can also reduce robustness.

Unlike statistical models, deformable segmentation models do not necessarily require training data or prior knowledge. Their capacity to deform allows them to adapt flexibly to target structures. However, they are sensitive to contour initialization and stopping criteria, both of which depend on image-specific characteristics; this sensitivity limits generalization across large image collections.

Machine-learning methods also learn predictive models from training data, but they estimate a class probability for each voxel. These probabilities can be used as unary potentials in regularized image-segmentation formulations, enabling voxel classification and segmentation. Traditional machine-learning pipelines separate handcrafted feature extraction from classification. Although such pipelines have performed well on some tasks, classifier design can be laborious, and domain experts must select both the task-specific features and classifier hyperparameters. These manual choices may not be optimal and restrict applicability across anatomical structures and imaging modalities. Deep learning, particularly convolutional neural networks, instead learns task-relevant representations end to end without manually specifying features or parameters. The distinction between an atlas-based workflow and an end-to-end deep-learning workflow is illustrated in Figure~\ref{fig:segmentation-paradigms}.

Deep-learning methods have therefore attracted considerable attention in medical image segmentation.

\subsection{Deep Learning-Based Semantic Segmentation}

Deep learning refers to multilayer neural networks that extract hierarchically organized features from raw inputs. In contrast to handcrafted feature extraction in conventional machine learning, deep networks learn complex feature hierarchies automatically. Accurate prediction is learned from large datasets, and rapid advances in graphics processing units have enabled deep-learning algorithms to flourish across many tasks. The convolutional neural network (CNN), first proposed in 1989\cite{ref28}, became particularly prominent after its success in the 2012 ImageNet challenge\cite{ref29,ref30,ref31}.

\begin{figure}[tbp]
\centering
\includegraphics[width=0.86\linewidth]{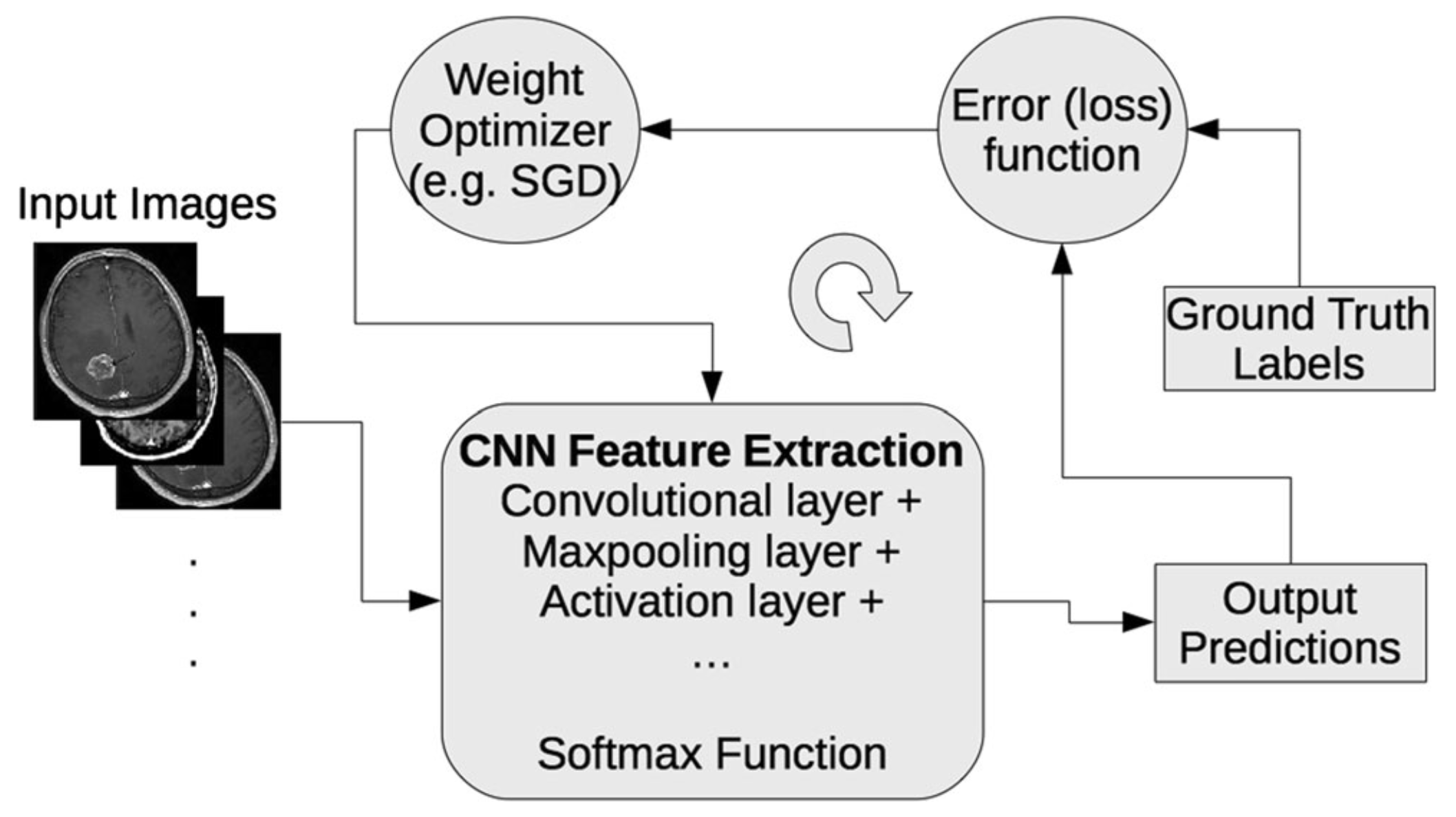}
\caption{Forward prediction and backpropagation in a convolutional neural network.}
\label{fig:cnn-training}
\end{figure}

A typical CNN contains convolution, pooling, activation, and batch-normalization layers. Modern networks may contain tens or hundreds of layers, millions of weights, and billions of connections. Convolutional layers filter an input image or feature map to generate new feature maps. Pooling layers downsample the preceding convolutional output by propagating the maximum or mean value within a defined neighborhood. Activation layers provide nonlinear feature mappings, commonly using the rectified linear unit (ReLU) or variants such as PReLU and leaky ReLU. Optional batch-normalization layers constrain the optimization space, accelerate convergence, and help reduce overfitting. Because CNNs have complex parameterizations, they generally require substantial labeled data for supervised training. During training, the output probability distribution is connected to a loss function, and the parameters are adjusted to minimize the discrepancy between the predictions and labels under regularization constraints. Backpropagation through the computational graph updates the weights at each iteration until convergence, as shown in Figure~\ref{fig:cnn-training}.

Computer-vision tasks for object recognition and scene understanding include image classification, object detection, semantic segmentation, instance segmentation, and panoptic segmentation. Image classification maps an input image to a one-dimensional probability distribution and assigns an image-level label. Object detection extends this formulation from global to local image understanding by localizing objects with bounding boxes and assigning a class to each box. Semantic segmentation further delineates object boundaries and assigns a class to every pixel. A deep network extracts contextual features, interprets image content, and partitions the image into regions corresponding to different object classes. Instance segmentation combines detection and segmentation: it first localizes and classifies each object and then predicts a binary mask for the pixels within each box. Panoptic segmentation generalizes this idea to the entire image by assigning both a semantic class and an instance identity to each pixel. For example, image classification can identify a cat in an image; object detection can localize and distinguish cats and dogs, including multiple animals of the same class; semantic segmentation labels all cat pixels as one class and all dog pixels as another; instance segmentation delineates each detected animal within its box; and panoptic segmentation partitions the full image while assigning different instance identifiers to individual cats.

General-purpose classification CNNs such as VGGNet\cite{ref32} commonly end with fully connected layers that estimate class probabilities. These outputs are one-dimensional and therefore unsuitable for dense semantic segmentation. Long et al. proposed the fully convolutional network (FCN)\cite{ref10}, in which the final fully connected layers are replaced by convolutions. The resulting two- or three-dimensional feature maps are followed by softmax classification to assign a class to each spatial location.

\subsubsection{Fully Convolutional Networks}

\begin{figure}[!htbp]
\centering
\begin{subfigure}{0.92\linewidth}
\centering\includegraphics[width=\linewidth]{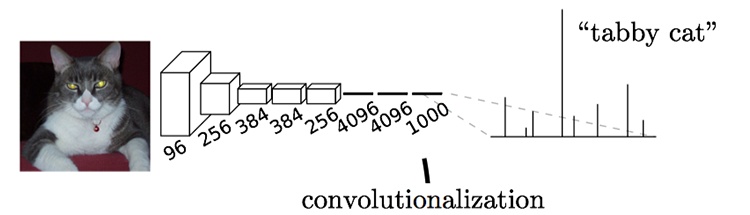}
\caption{Conventional CNN classification.}
\end{subfigure}
\par\medskip
\begin{subfigure}{0.92\linewidth}
\centering\includegraphics[width=\linewidth]{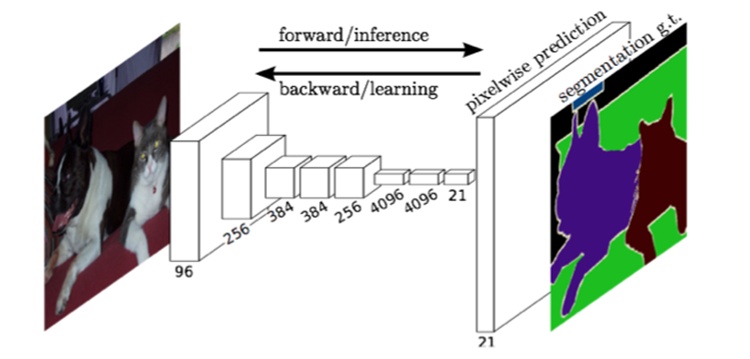}
\caption{FCN architecture.}
\end{subfigure}
\par\medskip
\begin{subfigure}{0.86\linewidth}
\centering\includegraphics[width=\linewidth]{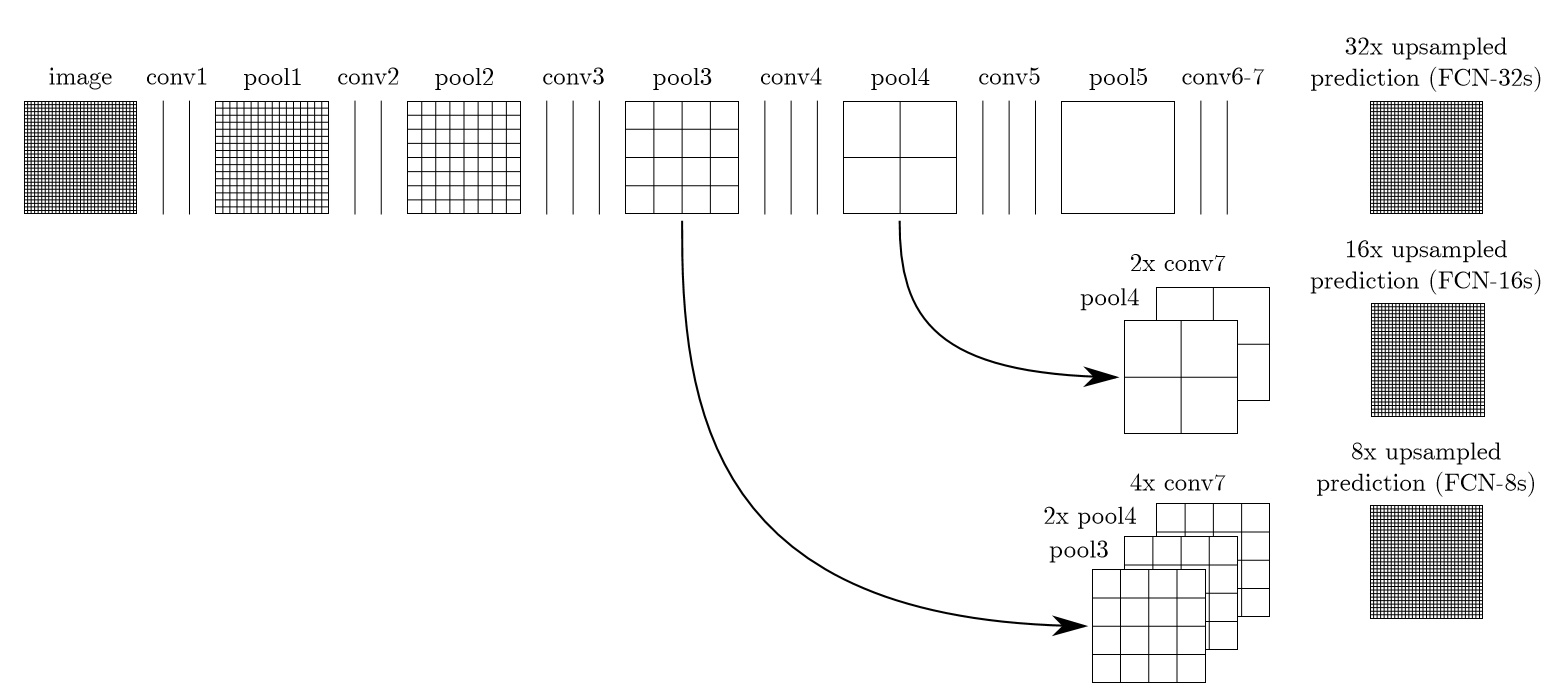}
\caption{Downsampling and upsampling in an FCN.}
\end{subfigure}
\caption{Comparison of CNN and FCN architectures.}
\label{fig:cnn-fcn}
\end{figure}

Since 2015, FCNs derived from CNNs have become a principal approach to semantic segmentation because they accept arbitrary input sizes, perform dense prediction, and support diverse architectures with good generalization\cite{ref10}. An important feature of FCNs is that existing CNNs can be reused as modules for hierarchical feature extraction. A conventional CNN vectorizes the feature maps after feature extraction and applies fully connected layers to transform and integrate the features into a fixed-length vector. It therefore predicts a probability distribution for an entire image and is appropriate for image-level classification or regression. AlexNet trained on ImageNet, for example, outputs a 1,000-dimensional vector whose softmax-normalized entries represent class probabilities. In Figure~\ref{fig:cnn-fcn}a, an image of a cat is supplied to AlexNet and the class ``tabby cat'' receives the highest probability.

An FCN instead classifies every pixel. Whereas a conventional CNN uses a fully connected network over one-dimensional neurons for integration and classification, an FCN uses $1\times1$ convolutions over multidimensional feature maps. It can accept an arbitrary input size and commonly upsamples the final convolutional feature map to the input resolution. A class probability distribution is then generated at every spatial location. During training, the softmax loss is computed at the pixel level, with each pixel serving as a training sample.

As shown in Figure~\ref{fig:cnn-fcn}b, an FCN first downsamples an image using convolution and pooling and then restores spatial resolution using transposed convolution. Downsampling extracts semantic information but produces small feature maps, which must be returned to the original image size. In the downsampling portion shown in Figure~\ref{fig:cnn-fcn}c, several convolutions and a max-pooling layer transform the input into the \texttt{pool1} feature maps at half the original width and height. Repetition produces \texttt{pool2} at one-quarter resolution, \texttt{pool3} at one-eighth resolution, and ultimately \texttt{pool5} at one-thirty-second resolution. FCN-32s upsamples \texttt{pool5} directly by a factor of 32 and applies softmax to obtain the segmentation. FCN-16s first upsamples \texttt{pool5} by a factor of two, adds the result pointwise to \texttt{pool4}, and upsamples the fused feature map by a factor of 16. FCN-8s adds a further fusion with \texttt{pool3}. Units consisting of multiple convolutions and a pooling layer can be derived from established convolutional classification networks.

FCNs offer three principal advantages. First, existing classification networks can be converted into fully convolutional models by replacing fully connected layers with convolutional layers, producing spatial maps rather than one-dimensional scores. Second, they establish an end-to-end training formulation for semantic segmentation. Third, they efficiently learn pixel-level predictions from inputs of arbitrary size. Many subsequent segmentation methods inherited and extended the FCN formulation. The four FCN-based comparator networks used in this thesis---UNet, VoxResNet, DeepMedic, and HyperDense-Net---are introduced below.

\subsubsection{UNet}

UNet was introduced at MICCAI 2015\cite{ref11}, followed by its three-dimensional version at MICCAI 2016\cite{ref12}. It became one of the most widely adopted semantic-segmentation architectures and has been applied to natural and medical images in both two and three dimensions. Like the original FCN, UNet follows an encoder--decoder design. A symmetric upsampling pathway maps encoded features back into the output space, while parallel skip connections fuse feature distributions across scales and improve generalization. UNet subsequently became a standard baseline for medical image segmentation and inspired many U-shaped architectures. Related U-shaped designs were also adopted in methods such as VoxResNet\cite{ref13}, the discriminative feature network\cite{ref33}, and the feature pyramid network\cite{ref34}.

\begin{figure}[tbp]
\centering
\includegraphics[width=0.82\linewidth]{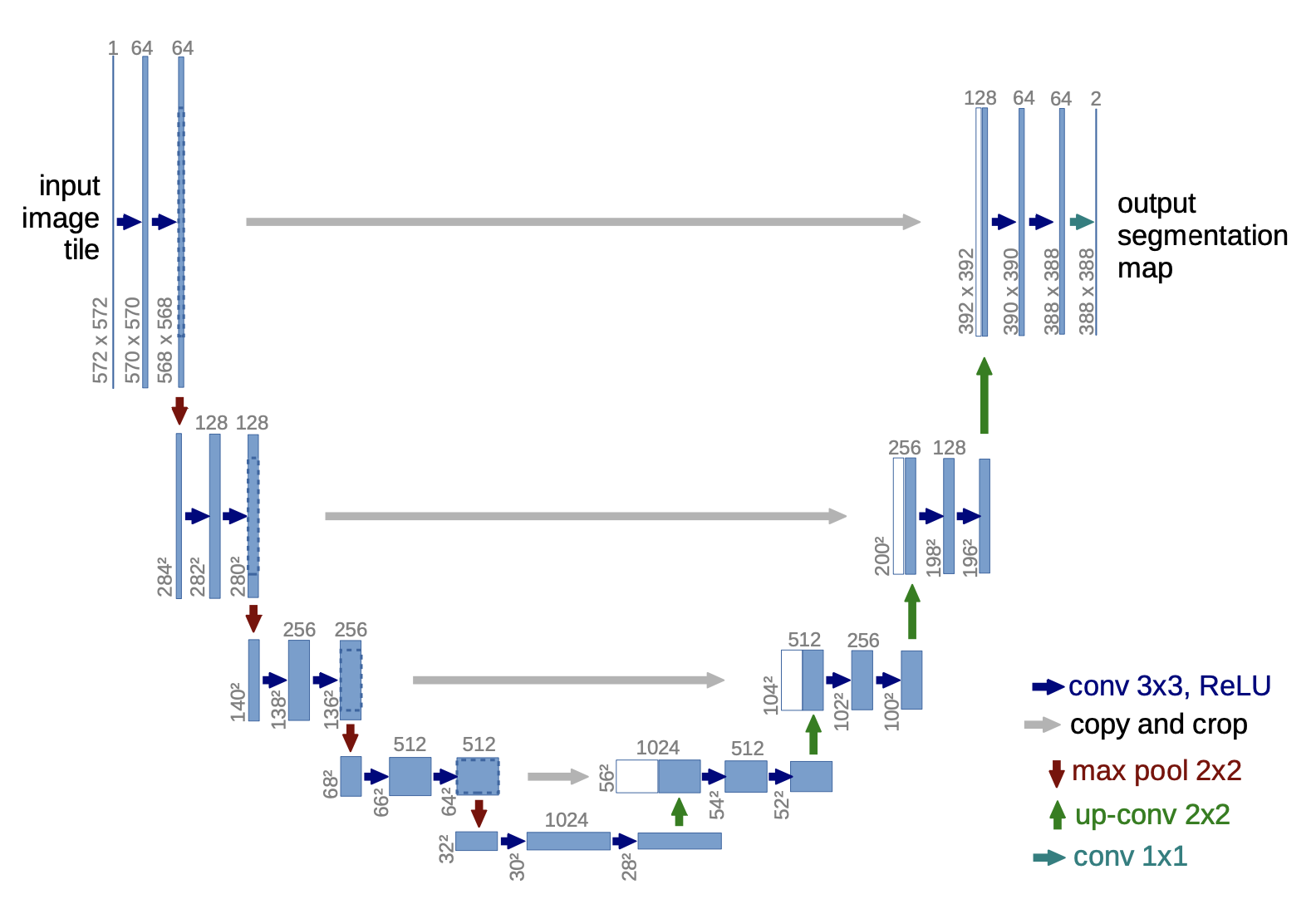}
\caption{Schematic architecture of UNet.}
\label{fig:unet}
\end{figure}

The two defining features of UNet are its U-shaped architecture and parallel skip connections (Figure~\ref{fig:unet}). The encoder contains four stages, each with two convolutional layers followed by max pooling that reduces the feature-map resolution by a factor of two. The decoder likewise contains four stages and progressively upsamples the feature maps to the input resolution. Parallel skip connections join corresponding encoder and decoder stages, allowing shallow local information to compensate for local detail lost from deeper global representations during upsampling. The original FCN uses transposed convolution for upsampling, whereas the implementation described here uses interpolation in UNet. The FCN fuses shallow and deep features by pointwise addition, whereas UNet concatenates them along the channel dimension. Addition incorporates information without increasing dimensionality but may alter the interpolated feature distribution. Concatenation preserves more channel and spatial information and allows later convolutions to select freely between shallow and deep features.

Although UNet established the U-shaped FCN family and achieved strong performance, it has several limitations in medical image segmentation. First, skip connections compensate for information loss during upsampling, but high-level feature maps can still lack texture, reducing sensitivity to object boundaries. Second, the large receptive field created by encoding and decoding is advantageous for large targets but may be less sensitive to small targets. Third, a single symmetric U-shaped structure is not readily extended to multiple pathways for multimodal imaging. Fourth, three-dimensional UNet uses substantial memory, leaving limited capacity for further architectural expansion. Subsequent work reduced parameter counts and improved segmentation through architecture and loss-function design\cite{ref13}, while multiscale receptive fields were used to accommodate anatomical structures of different sizes and shapes. From 2018 onward, many studies also shifted attention from architecture design to feature selection, introducing mechanisms such as channel attention\cite{ref14}, context encoding\cite{ref15}, and nonlocal self-attention\cite{ref16}. Such mechanisms could also be placed above a new backbone that addresses boundary and small-structure sensitivity at the architectural level.

\subsubsection{VoxResNet}

UNet provides a strong baseline, but its parameter count limits extensibility in three-dimensional segmentation. Information can also be lost during downsampling and upsampling, and the relatively fixed receptive field may impair segmentation of small anatomical structures. Chen et al. proposed VoxResNet in 2017\cite{ref13}. Residual units strengthen semantic representations while enabling a lighter architecture. Multiple upsampling pathways also support deep supervision: output maps from several layers are included in the training objective, providing multiscale features and variable receptive fields for structures of different sizes.

\begin{figure}[tbp]
\centering
\includegraphics[width=0.96\linewidth]{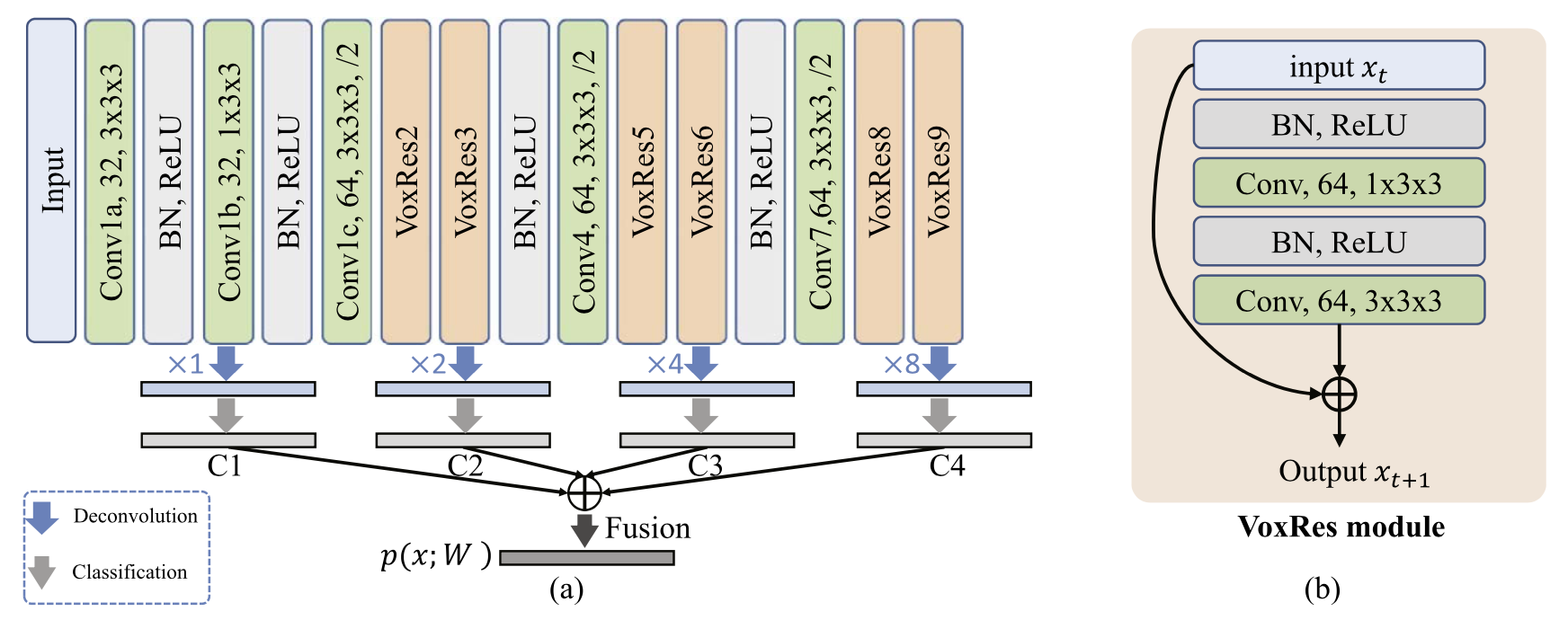}
\caption{Schematic architecture of VoxResNet.}
\label{fig:voxresnet}
\end{figure}

VoxResNet contains one downsampling pathway and four upsampling pathways at different scales (Figure~\ref{fig:voxresnet}). During inference, same-sized feature maps produced by the four pathways are summed to give $C=C_1+C_2+C_3+C_4$. During training, the segmentation accuracy of $C_1$, $C_2$, and $C_3$ is included in the objective in addition to the final output $C$. Thus, the model learns multiscale features under different receptive fields. \texttt{Conv1c}, \texttt{Conv4}, and \texttt{Conv7} use stride 2. The four blue arrows at each scale denote transposed convolutions with strides 1, 2, 4, and 8; the gray arrows denote classification. The resulting maps $C_1$ through $C_4$ are summed to form the final segmentation-probability map. Batch normalization (BN) and ReLU activation are used.

VoxResNet uses the following deeply supervised objective:
\begin{equation}
\mathcal{L}(x,y;\theta)=\lambda\psi(\theta)
-\sum_{\alpha}\sum_{x\in\mathcal{V}}\sum_c
\omega_{\alpha}y_c^x\log p_c^{\alpha}(x;\theta)
-\sum_{x\in\mathcal{V}}\sum_c y_c^x\log p_c(x;\theta).
\end{equation}
The first term is $L_2$ regularization, weighted by $\lambda$. In the second term, the softmax-normalized outputs $p_c^{\alpha}(x;\theta)$ from intermediate maps act as auxiliary classifiers indexed by $\alpha$. In the third term, $p_c(x;\theta)$ is the final classifier obtained after softmax normalization of $C$, and $c$ indexes the class. The auxiliary weights $\omega_{\alpha}$ are initialized to 1 and decay to a small value such as $10^{-3}$. Deep supervision introduces receptive fields at different scales during training to accommodate anatomical structures with variable sizes and shapes.

Rather than placing residual connections only within a single path, VoxResNet uses several upsampling paths and combines them into one segmentation result. Auxiliary classifiers incorporate outputs from different scales and paths into the objective. Compared with UNet, this design is lightweight and performs well in segmentation of regional structures such as subcortical tissues and tree-like structures such as vessels. Its efficiency nevertheless depends on relatively small output-channel counts. If these widths are increased, four upsampling pathways can multiply GPU memory requirements. In three-dimensional medical segmentation, limited memory and large volumetric patches may favor omitting transposed convolution and mapping a selected feature-map scale directly to a segmentation, an approach shown to be effective in previous work\cite{ref17,ref18,ref19}.

\subsubsection{DeepMedic}

U-shaped FCNs restore feature-map resolution while extracting features and constitute a standard dense-inference architecture. Their sensitivity to boundaries and small structures can nevertheless be improved. Attention mechanisms\cite{ref14,ref15,ref16} use feature selection to improve this sensitivity and have been widely studied. A backbone designed specifically to address these limitations could also provide a stronger basis for such higher-level mechanisms. DeepMedic, proposed in 2017\cite{ref17}, and its application to subcortical segmentation in LiviaNET in 2018\cite{ref18}, motivated the present work. These approaches suggest that a lightweight and extensible architecture can improve perception of boundaries and small targets while incorporating multimodal context.

Whole-volume dense prediction is often impractical for three-dimensional medical images because of GPU memory requirements, so patch-based segmentation is commonly used. In conventional patch-wise segmentation, the input patch matches the receptive field of the final neuron, and the single output probability predicts the class of the central voxel. This approach avoids the memory cost of upsampling and expanding feature maps. DeepMedic replaces fully connected layers with convolutions and accepts an input patch larger than the receptive field, enabling dense prediction for several adjacent voxels at once. It therefore provides an intermediate strategy between point inference and whole-image dense inference. Unlike a U-shaped model, its input and output patch sizes differ because the input is larger than the receptive field. Dense outputs are obtained without a large memory burden. DeepMedic thus adapts the traditional patch-wise formulation to an FCN, omits the common upsampling path, and maps a coupled feature representation directly to segmentation probabilities through a lightweight downsampling-only network.

DeepMedic was initially used for brain-lesion segmentation. It introduced dense patch-based training and inference to reduce three-dimensional computation and parallel convolutional pathways to process multimodal information. This multiscale design balances receptive field against feature-map resolution. Equal-probability sampling from foreground and background regions also addresses class imbalance. Because patches retain an intrinsic sample distribution, the resulting predictions can balance sensitivity and specificity.

\begin{figure}[tbp]
\centering
\includegraphics[width=0.96\linewidth]{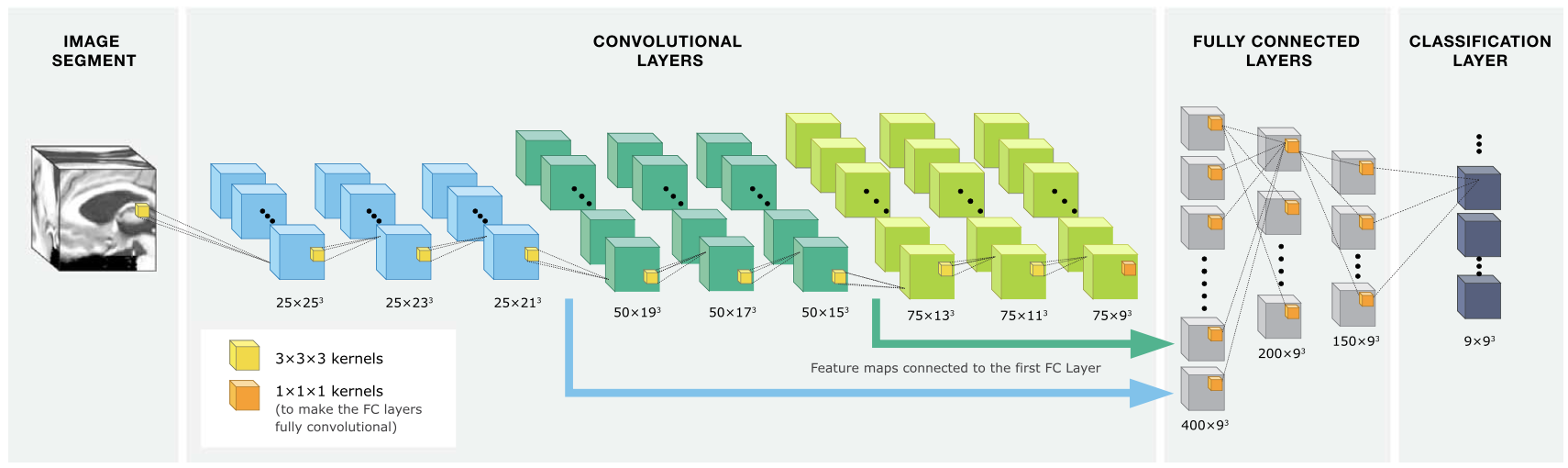}
\caption{Schematic architecture of DeepMedic.}
\label{fig:deepmedic}
\end{figure}

Dolz et al. subsequently applied the DeepMedic-style architecture LiviaNET to subcortical segmentation\cite{ref18}, as illustrated in Figure~\ref{fig:deepmedic}. Without an upsampling pathway, the output dimensions decrease approximately linearly with network depth. No padding is applied at feature-map boundaries; hence, each convolution reduces the spatial dimensions. If a kernel has size $k$ and the current feature dimension is $v$, the output dimension is
\begin{equation}
v'=v-k+1.
\end{equation}
The accumulation of small reductions allows the same sequence of convolutions to perform both downsampling and feature extraction. DeepMedic improves segmentation of small subcortical structures, but memory constraints limit its receptive field\cite{ref18}. It was used for brain-lesion segmentation by Kamnitsas et al.\cite{ref17} and for the thalamus, caudate nucleus, putamen, and pallidum by Dolz et al.\cite{ref18}. At the time of the thesis, it had not been applied to simultaneous whole-brain segmentation of both large and small subcortical structures. Expanding the receptive field without increasing the memory burden therefore became a central design objective.

\subsubsection{HyperDense-Net}

HyperDense-Net extends the downsampling-only FCN family\cite{ref19}. It inherits DeepMedic's downsampling scheme and dense prediction over patches larger than the receptive field. Dense connections propagate each layer's feature distribution to all subsequent feature spaces, encourage feature reuse, facilitate depth, and mitigate vanishing gradients, improving multimodal segmentation. The method focuses on multimodal images, however, and does not provide a consistent improvement for single-modal segmentation. DeepMedic has therefore often been used for single-modal images and HyperDense-Net for multimodal images, leaving no shared framework that improves on existing methods for both settings.

\begin{figure}[!htbp]
\centering
\begin{subfigure}{0.96\linewidth}
\centering\includegraphics[width=\linewidth]{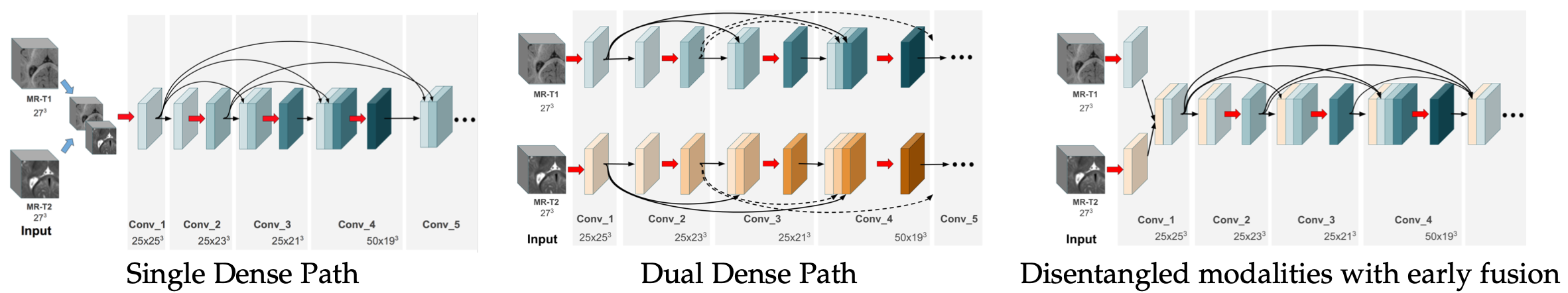}
\caption{Three multimodal fusion strategies.}
\end{subfigure}
\par\medskip
\begin{subfigure}{0.76\linewidth}
\centering\includegraphics[width=\linewidth]{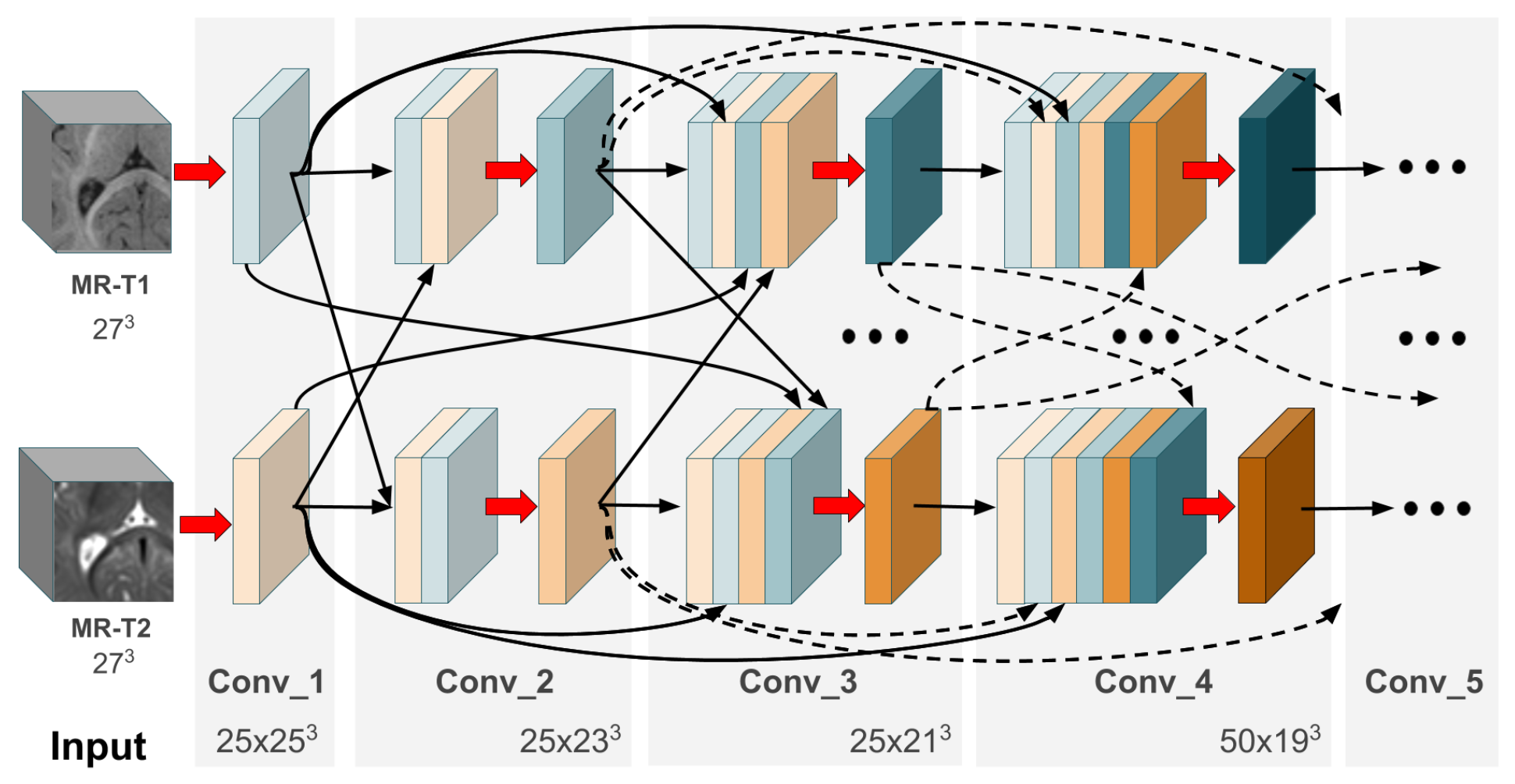}
\caption{HyperDense-Net and its cross-path fusion.}
\end{subfigure}
\caption{Comparison of multimodal information-fusion strategies.}
\label{fig:multimodal-fusion}
\end{figure}

Within each path, HyperDense-Net connects the output of every layer to the inputs of all subsequent layers. Three conventional multimodal fusion strategies are illustrated in Figure~\ref{fig:multimodal-fusion}a. A single-path strategy concatenates the original multimodal patches and extracts their low-level features jointly from the first convolution onward. A multipath strategy estimates each modality independently and fuses the paths only at the end. An early-fusion disentangled-modality strategy extracts low-level features using separate first-layer convolutions, concatenates them, and then estimates high-level abstract features in one path.

Permanent fusion within one path produces modality-unbiased feature distributions and makes it difficult to retain modality-specific semantic information. Independent paths can estimate modality-specific distributions but exchange no semantic information before the final fusion, limiting their ability to combine shared multimodal information. Because different modalities image the same anatomical region despite differing in intensity distribution, information must flow between paths while each path estimates modality-specific features. The early-fusion disentangled strategy therefore extracts biased local features separately at low levels and merges them into a single path that estimates an unbiased high-level feature distribution.

HyperDense-Net balances modality-specific and modality-unbiased feature estimation and jointly extracts semantic information shared across modalities as well as information specific to each modality. It uses two dense paths, one per modality (Figure~\ref{fig:multimodal-fusion}b). Dense connections propagate features both to all later layers in the same path and to all later layers in the other path. Each convolution can therefore estimate features from its own modality while incorporating co-located features from the other modality. For single-modal segmentation, however, the same modality must be supplied to multiple paths. This increases the parameter count without introducing complementary modality distributions; consequently, the architecture is less well suited to single-modal tasks. DeepMedic and HyperDense-Net therefore did not provide one effective architecture across both modality settings.

\section{Research Objectives and Scope}

Motivated by the preceding background, this thesis studies downsampling FCNs based on deep learning and patch-wise dense inference for segmenting subcortical structures in MR images. The principal research questions are:
\begin{enumerate}
\item how to revise feature-map downsampling in a downsampling FCN so that the receptive field is enlarged without a substantial increase in parameters;
\item how to improve segmentation while reducing the number of network parameters under the dataset-size and memory constraints of medical image segmentation; and
\item how to design one network that unifies single- and multimodal segmentation, for which separate state-of-the-art designs previously existed.
\end{enumerate}

The corresponding objectives are:
\begin{enumerate}
\item to obtain a larger feature receptive field under the input--output relationship of DeepMedic-like downsampling FCNs without changing the order of parameter magnitude;
\item to design a better multiscale information-fusion strategy for single-modal tasks that improves segmentation while reducing model size; and
\item to extend the resulting single-modal architecture to multimodal tasks while retaining competitive performance.
\end{enumerate}

The study has both methodological and potential clinical significance. With a fixed memory budget, a larger receptive field can reduce information loss from the input and improve downsampling-FCN accuracy across structures of different scales. Medical imaging datasets typically contain relatively few images, each with a large volume. Improving accuracy while reducing the number of parameters may reduce overfitting on small datasets and lower memory requirements. A common architecture for single- and dual-modal tasks may also reduce the time required to select and tune task-specific networks. Clinically, rapid and accurate segmentation can support precise measurement of indicators relevant to neurological disease and reduce the burden of manual voxel-level delineation, allowing clinicians to devote more time to direct care and communication with patients.

Under these objectives, the thesis proposes DenseMedic and ACNN for rapid automatic segmentation of subcortical structures in brain MR images. DenseMedic addresses single-modal T1-weighted MR images, whereas ACNN addresses both single- and multimodal MR images. DenseMedic introduces OreoDown, an abstract design framework for downsampling FCNs that explicitly separates size-changing downsampling from size-preserving feature extraction. It instantiates downsampling with rapid spatial reduction and feature extraction with densely connected layers, and then employs hybrid dilated convolutions to refine feature perception over a large receptive field. Within the OreoDown framework, ACNN instead instantiates feature extraction using alternate-connected layers. Before the addition of hybrid dilated convolution, its stated parameter count is approximately one third that of DenseMedic while its segmentation performance improves. Dividing the single path centrally creates a dual-path ACNN without changing the stated parameter count; this model is applied to single-modal T1 and dual-modal T1+FLAIR and T1+T1-IR tasks. Finally, prior information is introduced for single-modal segmentation so that T1-weighted inputs with slightly different intensity distributions are supplied to the two paths.

\section{Organization of the Thesis}

\begin{figure}[!htbp]
\centering
\includegraphics[width=0.82\linewidth]{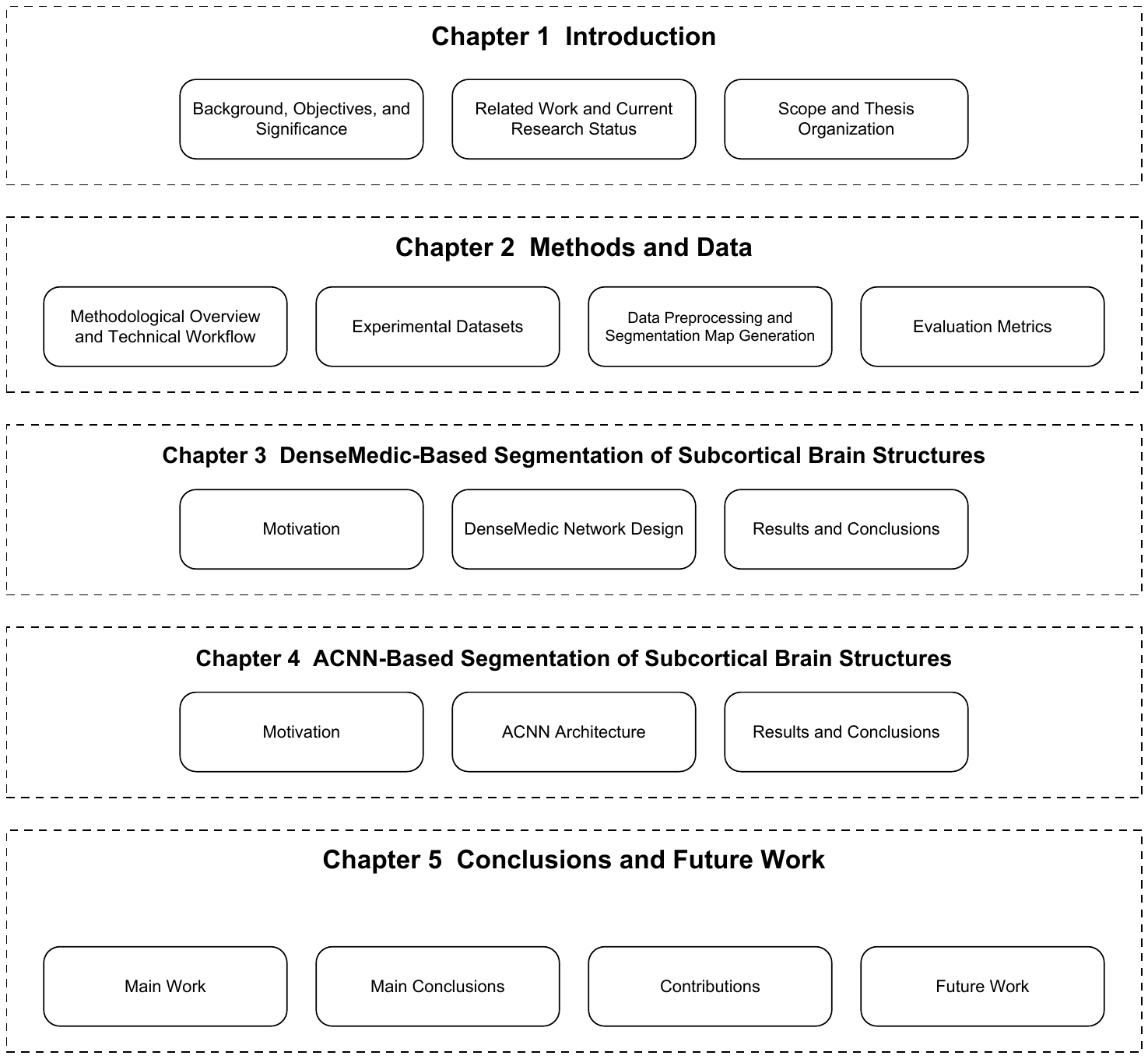}
\caption{Overall organization of the thesis.}
\label{fig:thesis-organization}
\end{figure}

The organization of the thesis is summarized in Figure~\ref{fig:thesis-organization}.

Chapter 1 introduces the background, purpose, and significance of the study. It reviews atlas-based, statistical, deformable, and machine-learning methods for medical image segmentation and describes the FCN, UNet, VoxResNet, DeepMedic, and HyperDense-Net architectures. The research objectives, scope, and principal contributions are then presented.

Chapter 2 describes the experimental data and methodology. It presents the technical workflow, introduces the IBSR and MRBrainS18 datasets, describes preprocessing, patch sampling, segmentation-map reconstruction, and cross-validation, and defines the evaluation metrics and computing environment.

Chapter 3 presents DenseMedic for subcortical segmentation. It introduces the motivation and three main design components: the OreoDown framework, which separates downsampling from feature extraction; the instantiation of OreoDown using rapid downsampling and densely connected feature-extraction layers; and hybrid dilated convolutions for refined perception over a large receptive field. Results on both datasets and a discussion are provided.

Chapter 4 presents ACNN. It introduces alternate connectivity, the division of a single-path network into a dual-path network, and the incorporation of prior information for single-modal segmentation. Results on single- and dual-modal tasks are analyzed and discussed.

Chapter 5 summarizes the methods, conclusions, contributions, and directions for future work.

\chapter{Overview of the Methods and Data}

\section{Technical Workflow}

For single- or multimodal magnetic resonance (MR) images, this study adopts a fully convolutional network (FCN)-based semantic-segmentation method. On the basis of dense inference and dense training, the images are segmented patch by patch and a class label is assigned to every voxel. Because supervised deep learning is used, a complete segmentation procedure for a neural network consists principally of three stages: data preprocessing, training, and inference. During preprocessing, the network input $x$ and its corresponding reference segmentation $y$ must be processed and sampled. During training, iterative optimization is performed on labeled data so that the network output $y'$ for input $x$ increasingly approximates the reference segmentation $y$. After convergence, the inference stage applies the trained network to a test MR image to produce its corresponding segmentation map.

\begin{figure}[!t]
\centering
\includegraphics[height=0.82\textheight,keepaspectratio]{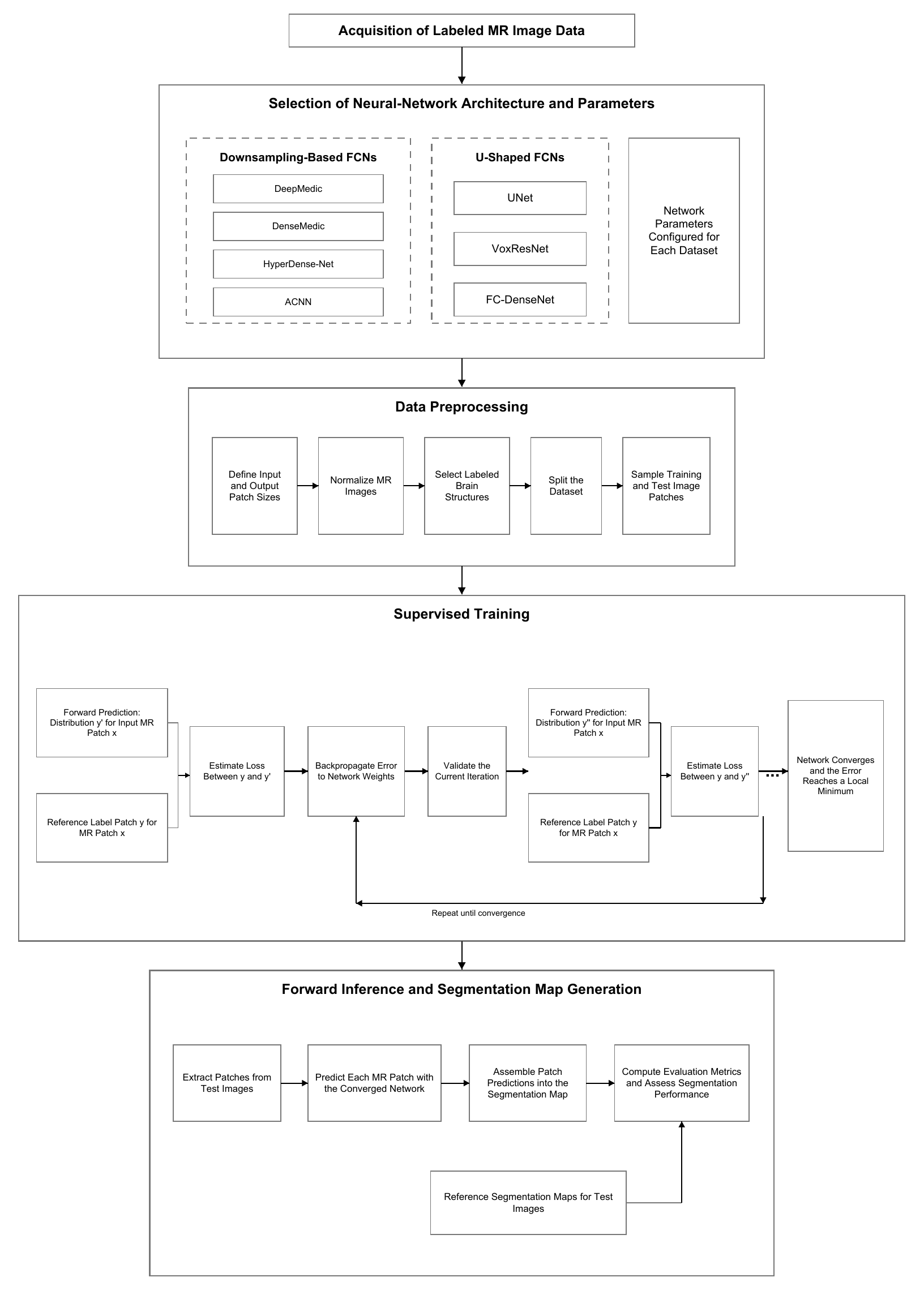}
\caption{Technical workflow of the FCN-based semantic-segmentation method.}
\label{fig:workflow}
\end{figure}

The technical workflow is shown in Figure~\ref{fig:workflow}. A complete procedure contains five components: acquisition of labeled MR images, network architecture design and parameter selection, data preprocessing, supervised training, and segmentation-map generation. In the first stage, a sufficient number of annotated MR images are selected for training and validation. Test MR images with a similar intensity distribution and their annotations are also specified; these annotations are used to evaluate the segmentations generated by the converged network. In the second stage, the network architecture and its hyperparameters are selected according to the characteristics of the dataset acquired in the first stage. Common FCN-based semantic-segmentation architectures include U-shaped FCNs and the downsampling FCNs developed further in this thesis.

Because medical-image segmentation commonly operates on patches, determining the network architecture also determines the input- and output-patch sizes required in the third, preprocessing, stage. The same intensity-normalization procedure is then applied to all MR images, and the same class-selection procedure is applied to all corresponding reference annotations. After this image processing, cases are divided into training and validation sets, with a fixed test set held out. The training and validation sets may be randomly partitioned several times at a predefined ratio. MR-image patches and corresponding reference-label patches are then sampled from the training and validation cases to produce labeled data for network training.

In the fourth stage, these labeled data are used for supervised training. The network parameters are initialized first. During each iteration, a forward pass produces the probability-distribution prediction $y'$ for every training patch $x$, and a loss function compares this prediction with the corresponding reference patch $y$. The error is then backpropagated through the network, and stochastic gradient descent or a related method adjusts each weight so that the error decreases as rapidly as possible toward a local minimum. After the update for the current iteration, a forward pass on the validation data provides an accuracy or error measurement that helps prevent excessive fitting to the training set. After multiple iterations, the network converges and the training error approaches a local minimum, completing supervised training.

In the fifth and final stage, the converged network segments each test image and its performance is evaluated by computing metrics between the resulting segmentation and the corresponding reference annotation. During segmentation, a test MR image is first divided along a regular three-dimensional grid to obtain the original input patches. Forward inference produces a segmentation for every input patch, and these output patches are then assembled in reverse order to reconstruct the segmentation map of the MR image. The resulting maps are used to assess segmentation performance.

Against this workflow, Section~\ref{sec:datasets} introduces the annotated MR datasets acquired in the first stage. Chapters~3 and~4 describe the architectures of DenseMedic and ACNN, respectively, together with the architectural hyperparameters and corresponding input- and output-patch sizes. Section~\ref{sec:preprocessing} describes structure selection, dataset partitioning, and sampling of training and validation patches. With the supervised-training procedure established, Sections~3.3.1 and~4.3.1 present implementation details such as the loss function, optimizer, and learning-rate settings. Finally, Sections~3.3 and~3.4 report and analyze the segmentation results of DenseMedic, while Section~4.3 reports and analyzes those of ACNN.

\section{Experimental Datasets}
\label{sec:datasets}

\subsection{IBSR}

The Internet Brain Segmentation Repository (IBSR; \url{https://www.nitrc.org/projects/ibsr})\cite{ref35} is a public dataset commonly used for subcortical brain segmentation, and numerous published studies have used it to evaluate algorithm performance. The dataset contains 18 T1-weighted MR (T1-MR) images from 18 subjects and their corresponding ground-truth annotations. Each volume has dimensions of $256\times128\times256$ voxels and is stored in NIfTI format. In each ground-truth annotation, experts delineated 36 lateralized subcortical structures, as listed in Table~\ref{tab:original-datasets}.

\subsection{MRBrainS18}

The MRBrainS18 dataset originates from the Grand Challenge on MR Brain Segmentation held at MICCAI 2018 (\url{https://mrbrains18.isi.uu.nl}). The images were acquired using a 3-T MR scanner at Utrecht University in the Netherlands. The organizers released multimodal MR images from seven subjects for network training and validation. The modalities comprise T1, T1-weighted inversion recovery (T1-IR), and T2-weighted fluid-attenuated inversion recovery (T2-FLAIR), accompanied by the corresponding ground-truth annotations. The multimodal images and annotations had been calibrated and spatially aligned. The voxel dimensions were $0.958\,\mathrm{mm}\times0.958\,\mathrm{mm}\times3.0\,\mathrm{mm}$, and each image volume had dimensions of $240\times240\times48$ voxels. N4ITK bias-field correction\cite{ref36} had also been applied to the MR image of each modality.

Of the seven multimodal training cases in MRBrainS18, five (1, 4, 5, 7, and 14) originated from the earlier MRBrainS13 dataset (\url{https://mrbrains13.isi.uu.nl}); some inaccurate annotations had been corrected. The eight annotated subcortical structures or regions are listed in Table~\ref{tab:original-datasets}. The two additional MRBrainS18 training cases (070 and 148) also contain an infarct annotation. Because the other five cases do not contain this class, infarct regions were not segmented in this study. Among the eight remaining structures or regions, white-matter lesions require the T2-FLAIR modality, and accurate delineation of the outer cerebrospinal-fluid boundary requires both T1 and T1-IR; the remaining structures can be segmented from T1 alone.

\begin{table}[!htbp]
\centering
\caption{Original datasets for subcortical brain-structure segmentation.}
\label{tab:original-datasets}
\small
\setlength{\tabcolsep}{4pt}
\begin{tabularx}{\textwidth}{@{}p{0.15\textwidth}>{\centering\arraybackslash}p{0.09\textwidth}p{0.14\textwidth}>{\centering\arraybackslash}p{0.13\textwidth}X@{}}
\toprule
Dataset & Subjects & Case identifiers & Classes incl. background & Subcortical structures or regions \\
\midrule
IBSR & 18 & IBSR\_01--IBSR\_18 & 37 & Left and right cerebral white matter; left and right cerebral cortex; left and right cerebellar white matter; left and right cerebellar cortex; left and right lateral ventricles; left and right inferior lateral ventricles; left and right thalami; left and right caudate nuclei; left and right putamina; left and right pallida; third, fourth, and fifth ventricles; brain stem; left and right hippocampi; left and right amygdalae; cerebrospinal fluid; left and right accumbens areas; left and right ventral diencephalon; left and right choroid plexuses; optic chiasm. \\
\addlinespace
MRBrainS18 & 7 & 1, 4, 5, 7, 14, 070, 148 & 9 & Cortical gray matter; white matter; basal ganglia; white-matter lesions; cerebrospinal fluid; ventricles; cerebellum; brain stem. \\
\bottomrule
\end{tabularx}
\end{table}

\section{Image-Data Preprocessing}
\label{sec:preprocessing}

\subsection{Selection of Brain Structures}

For IBSR, the left- and right-hemisphere components of each tissue structure were merged, yielding 15 principal non-lateralized brain structures excluding the ventricles. Some small regions contain few voxels and therefore do not adequately preserve the randomness of the center-point sampling method of Wachinger et al.\cite{ref37} during training-data preprocessing. Consequently, 11 relatively large structures were selected from these 15 structures: brain stem (BS), hippocampus (HI), thalamus proper (TH), caudate (CA), putamen (PU), ventral diencephalon (VE), lateral ventricle (LV), cerebral white matter (WM), cerebral cortex (CT), cerebellar white matter (CWM), and cerebellar cortex (CCT). All remaining structures were merged into a separate ``other'' class. The four principal structures excluded from the 15 were the pallidum, amygdala, accumbens area, and choroid plexus.

For MRBrainS18, some subcortical structures or regions can be segmented from a single MR modality, whereas others require multimodal inputs. The T1-only task therefore segmented cortical gray matter, basal ganglia, white matter, ventricles, cerebellum, and brain stem. The dual-modal T1+T2-FLAIR task segmented cortical gray matter, basal ganglia, white matter, white-matter lesions, ventricles, cerebellum, and brain stem. The dual-modal T1+T1-IR task segmented cortical gray matter, basal ganglia, white matter, cerebrospinal fluid, ventricles, cerebellum, and brain stem. Apart from background, every unlisted structure was merged into an ``other'' class. The abbreviations used are cortical gray matter (CGM), basal ganglia (BG), white matter (WM), white-matter lesions (WML), cerebrospinal fluid in the extracerebral space (CRF), ventricle (VT), cerebellum (CB), and brain stem (BS).

\subsection{Training, Validation, and Test Partitions and Cross-Validation}

Cross-validation was used to evaluate segmentation performance. For cross-validation, each dataset was divided into a fixed test set and a fixed training--validation set that did not overlap. Holding the test data outside training provides a more effective evaluation of segmentation generalization. Keeping the test set fixed also permits controlled and reliable comparisons among networks. The fixed training--validation set was randomly divided at a specified ratio into a training set for supervised learning and a validation set for inference during each training iteration to prevent overfitting. Multiple random partitions of the training and validation data were generated; validation subsets generally did not overlap so that all data in the fixed training--validation set could contribute to performance validation.

Experiments on IBSR used three-fold cross-validation. The last two images in the dataset, IBSR\_17 and IBSR\_18, formed the fixed test set. The first 16 images were divided three times into training and validation subsets at a ratio of 14:2. Across these three partitions of the fixed training--validation set, the validation cases were IBSR\_01 and IBSR\_02, IBSR\_08 and IBSR\_09, and IBSR\_15 and IBSR\_16, respectively.

Experiments on MRBrainS18 likewise used three-fold cross-validation. For T1-only segmentation, case 14 was the fixed test case. Because this case belongs to the original MRBrainS13 dataset, the intention was to include as many of the healthier subjects from MRBrainS13 as possible in training; the validation case in the three random partitions was 070, 148, and 14, respectively. For dual-modal T1+T2-FLAIR segmentation, case 070 was the fixed test case and the validation cases in the three random partitions were 7, 14, and 148. For dual-modal T1+T1-IR segmentation, case 070 was the fixed test case and the three validation cases were 1, 4, and 148.

All networks evaluated for a given segmentation task used the same training, validation, and test partitions on both datasets.

\subsection{Training-Patch Sampling and Segmentation-Map Generation}

Because this study uses three-dimensional downsampling FCNs, let the input size of DenseMedic and ACNN be $v_0^3$ and the output size be $v_3^3$, where $v_0>v_3$. Depending on whether a task is single- or multimodal, preprocessing selects one or more MR images and the corresponding ground-truth annotation. The following describes Rules A and B in the data-processing stage of Figure~\ref{fig:workflow}.

\begin{figure}[tbp]
\centering
\includegraphics[width=\textwidth]{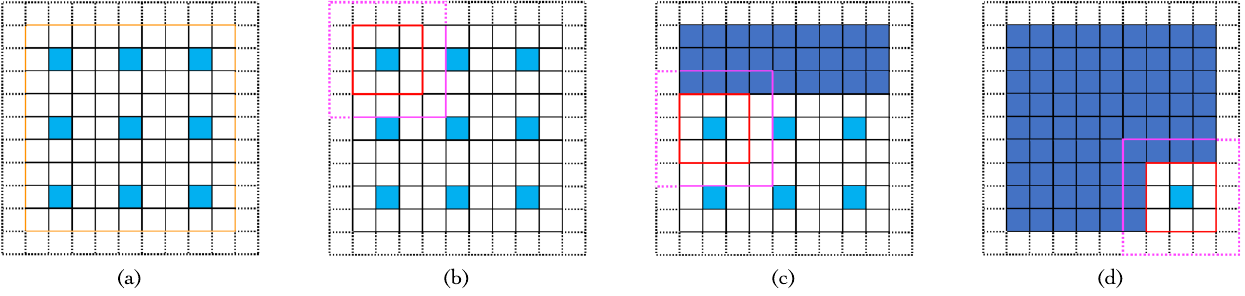}
\caption{Two-dimensional illustration of patch stacking during DenseMedic segmentation-map generation. The pink boxes denote input T1-MR patches ($v_0=5$), the red boxes denote output segmentation patches ($v_3=3$), light-blue cells denote patch centers, and dark-blue cells denote regions that have already been segmented. (a) Center-point sampling and image-boundary padding; (b) correspondence between input and output patches and segmentation around the first center point; (c) segmentation around the fourth center point; and (d) segmentation around the final center point.}
\label{fig:patch-stacking}
\end{figure}

For the training and validation sets, the patch sampling of Kamnitsas et al.\cite{ref9} was combined with the center-point sampling of Wachinger et al.\cite{ref37}. Patch sampling produces a fixed set of samples, whereas center-point sampling is random. For patch sampling, each three-dimensional ground-truth annotation in the training and validation sets was first divided into reference-label patches of size $v_3^3$. The coordinates of the center of each patch in the full annotation volume were retained. This procedure is the inverse of Rule B in Figure~\ref{fig:workflow}; the centers are shown as light-blue pixels in the two-dimensional illustration in Figure~\ref{fig:patch-stacking}a.

For any dimension whose original size was not divisible by $v_3$, the ground-truth annotation and its corresponding single or multiple MR images were padded with zeros in the axial direction until that dimension became divisible by $v_3$. The padded two-dimensional region lies inside the yellow box in Figure~\ref{fig:patch-stacking}a; the ground-truth and MR-image dimensions remained identical after this operation. At every retained center point, an MR patch of size $v_0^3$ was then extracted. A reference patch is shown by the red box in Figure~\ref{fig:patch-stacking}b, and its MR input patch is shown by the pink box. Multimodal tasks contain multiple co-located MR patches. Because $v_0>v_3$, each MR image was additionally padded with zeros by $(v_0-v_3)/2$ voxels along every boundary; this is the region outside the yellow box in Figure~\ref{fig:patch-stacking}a. These operations produced fixed samples for every image group. In Figure~\ref{fig:workflow}, obtaining $x$ by fixed sampling corresponds to Rule A.

Center-point sampling was then performed in addition to patch sampling. For each structure in a ground-truth annotation, center points were randomly selected until a total of 300 was reached. For example, if $h$ paired MR and reference patches obtained by fixed patch sampling had their centers in a given structure, a further $300-h$ centers were randomly selected among the voxels belonging to that structure. Because cerebral gray matter and cerebral white matter occupy relatively large volumes, 600 centers were selected in each of these two structures to reflect greater data diversity. A reference patch of size $v_3^3$ and an MR patch of size $v_0^3$ were then extracted around each center. If a patch extended beyond an image boundary, the corresponding boundary was zero-padded to permit extraction. This procedure generated random samples for every image group. Combining the fixed and random samples augmented the training and validation data. In Figure~\ref{fig:workflow}a, the resulting MR patches correspond to network input $x$, and the reference-label patches correspond to supervised target $y$.

During segmentation, DenseMedic and ACNN output segmentation patches of size $v_3^3$. These patches were assembled in reverse order into a full segmentation map. This reconstruction is the inverse of the procedure used to divide the ground-truth annotation during fixed patch sampling and corresponds to Rule B. Portions of the Rule B reconstruction process are shown in two dimensions in Figure~\ref{fig:patch-stacking}c and~d. In addition, each T1-MR image was divided by 255 before preprocessing to normalize its intensity values.

\section{Evaluation Metrics}

Following the cross-validation design, the fixed training--validation images in each dataset were partitioned three times at the corresponding ratio. The ratio was 14:2 for IBSR and 5:1 for MRBrainS18. The network produced by each of the three training runs segmented the fixed test set; consequently, every test image was segmented three times. Networks were evaluated from these repeated segmentations using four metrics. The Dice similarity coefficient (DSC) was defined as
\begin{equation}
\mathrm{DSC}=\frac{2\lvert S\cap R\rvert}{\lvert S\rvert+\lvert R\rvert}.
\end{equation}
The intersection over union (IoU) was
\begin{equation}
\mathrm{IoU}=\frac{\lvert S\cap R\rvert}{\lvert S\cup R\rvert}.
\end{equation}
The 95\% Hausdorff surface distance (HSD95) was
\begin{equation}
\mathrm{HSD95}=\frac{
\inf_{5\%,\,r\in R}d_m(r,\partial S)+
\inf_{5\%,\,s\in S}d_m(s,\partial R)}{2}.
\end{equation}
The average surface distance (ASD) was
\begin{equation}
\mathrm{ASD}=\frac{
\displaystyle\sum_{s\in\partial S}d_m(s,\partial R)+
\displaystyle\sum_{r\in\partial R}d_m(r,\partial S)}
{\lvert\partial S\rvert+\lvert\partial R\rvert}.
\end{equation}

For a given structure, $S$ denotes the set of its points in the predicted segmentation and $R$ the corresponding set in the ground-truth annotation; $\partial S$ and $\partial R$ are their respective boundary-point sets. Here, $\lvert V\rvert$ denotes the number of points in a set $V$, $d_m(v,V)$ is the minimum Euclidean distance between point $v$ and all points in $V$, and $\inf_{5\%,\,v\in V}(d_m)$ denotes the infimum of the largest 5\% of $d_m$ values in $V$.

All four metrics are commonly used to evaluate three-dimensional segmentation of subcortical brain structures. For every network, performance was summarized by the mean and standard deviation of each metric across the three segmentation results for all test images.

\section{Experimental Environment}

Networks were trained on a server using PyTorch 1.0.1 with Python 3.6.8 under Ubuntu 18.04.1. The server contained a 48-core CPU operating at 2.2\,GHz per core, 252\,GB of internal memory, two NVIDIA Tesla P100 GPUs with 12\,GB of memory each, and 10\,TB of external storage.

Inference was performed locally by the author using PyTorch 1.0.1 with Python 3.6.7 under Ubuntu 18.10. The local system contained a 16-core CPU operating at 2.1\,GHz per core, 62.8\,GB of internal memory, one NVIDIA GTX 1080 GPU with 8\,GB of memory, and 2\,TB of external storage.

\section{Chapter Summary}

This chapter introduced the overall methodology and data preparation. Section~2.1 outlined dense training and dense inference for FCN-based semantic segmentation and presented the workflows for training and prediction. Section~2.2 introduced the IBSR and MRBrainS18 datasets. Section~2.3 described preprocessing, including selection of label classes from the original datasets, partitioning into training, validation, and test data, cross-validation, sampling of training and validation patches, and reconstruction of segmentation maps. Section~2.4 defined four quantitative metrics for assessing segmentation performance, and Section~2.5 described the computing environments used for training and inference. With this workflow established, Chapters~3 and~4 design the specific network architectures used in the study.

\chapter{DenseMedic-Based Segmentation of Subcortical Brain Structures}

\section{Introduction}

FCN-based algorithms for subcortical brain-structure segmentation commonly use three-dimensional convolutions to capture interslice information more effectively. As noted in the description of DeepMedic in Section~1.2.2.4, one limitation of that architecture is its relatively small receptive field. For a three-dimensional FCN, the receptive field of the features output by layer $l$ with respect to the input patch is calculated as follows\cite{ref3,ref30}:
\begin{equation}
\boldsymbol{\varphi}_{l}=\boldsymbol{\varphi}_{l-1}+
\left(\boldsymbol{\kappa}_{l}-[1,1,1]\right)\circ
\prod_{i=1}^{l-1}\boldsymbol{\tau}_{i},
\label{eq:receptive-field}
\end{equation}
where $\boldsymbol{\varphi}_{l},\boldsymbol{\kappa}_{l},\boldsymbol{\tau}_{l}\in\mathbb{N}^{3}$ denote the receptive field, kernel size, and stride of layer $l$ along the three axes, respectively; $\prod_{i=1}^{l-1}\boldsymbol{\tau}_{i}$ is the cumulative coefficient; and $\circ$ denotes the Hadamard product.

On this basis, the receptive field of DeepMedic can be enlarged in three ways: (1) increasing the kernel size, (2) increasing network depth, or (3) increasing the stride. Their efficiency increases in that order when measured by the additional parameter count. For example, to double the receptive field of voxels in the segmentation output of DeepMedic\cite{ref13}, the first approach would enlarge every convolutional kernel from $[3,3,3]$ to $[5,5,5]$, increasing the number of parameters by $5^3/3^3=4.63$ times. The second approach would insert a size-preserving convolutional layer after every existing layer, doubling the parameter count. The third approach need only set the stride of the first layer to 2. The cumulative coefficient of every subsequent layer then doubles, such that $\prod_{i=1}^{l-1}\boldsymbol{\tau}_{i}=[2,2,2]$ for $l>1$, and size-preserving convolutions can restore the depth shortened by the larger stride. This achieves an approximately twofold receptive-field increase without adding parameters. Increasing the stride can therefore enlarge the receptive field of output voxels without additional memory consumption, addressing the restricted receptive field of DeepMedic. It also separates size-changing downsampling from size-preserving feature extraction, permitting the latter to be redesigned more appropriately whenever memory capacity allows.

UNet instead introduces a symmetric upsampling path, obtains output-space probability distributions through coupled encoding and decoding of the input space, and incorporates parallel skip connections in feature space to estimate multiscale distributions with strong generalization. Subsequent U-shaped networks further improved segmentation accuracy, including VoxResNet\cite{ref13}, which integrates the ResNet design of Chen et al.\cite{ref38}, and FC-DenseNet\cite{ref40,ref41}, which integrates DenseNet as proposed by J\'egou et al.\cite{ref39}. DenseNet propagates the feature distribution of the current layer forward into later feature spaces, facilitates deeper feature representations, encourages feature reuse, mitigates vanishing gradients, and promotes the expression of multiscale information. Accordingly, retaining the high accuracy of a downsampling FCN while redesigning its feature-extraction process with densely connected layers may further improve segmentation.

Dilated convolution\cite{ref42} is also widely used in semantic segmentation because it enlarges receptive fields directly and effectively\cite{ref43,ref44}. It inserts zeros between adjacent kernel parameters. For a kernel of size $k^3$, the effective dilated-kernel size is
\begin{equation}
k_d=k+(k-1)(r-1),
\end{equation}
where $r$ is the dilation rate; if $r>1$, then $k_d>k$. A dilated kernel occupying a volume of $k_d^3$, however, contains $k_d^3-k^3$ zero-valued positions. Features at these positions are completely ignored in the kernel output, making its feature sampling excessively coarse. A parallel set of dilated kernels with complementary dilation rates $r$ and effective sizes $k_d$ can cover all positions within the receptive field. Hybrid dilated convolution\cite{ref45} provides such a set: it expands the receptive field without a large parameter increase while addressing the intrinsically coarse sampling of an individual dilated kernel.

Based on this analysis of existing FCN-based brain-structure segmentation methods, this chapter proposes DenseMedic for rapid automatic segmentation of subcortical structures in brain MR images. Its segmentation procedure resembles that of DeepMedic but incorporates several designs intended to improve accuracy: sandwich-like downsampling, densely connected feature-extraction layers, and hybrid dilated convolutions. Together they provide rapid downsampling, stronger contextual perception in feature space, and a larger receptive field, thereby improving segmentation performance and robustness and potentially supporting rapid diagnosis and treatment of brain diseases.

\section{Design of DenseMedic}

The network consists of purple skeleton layers (SLs), red feature extractors (FEs), and green fully connected and classification (FC) layers. The skeleton is composed principally of downsamplers (DSs). Its first layer is a conventional convolution (Conv) that performs broad, size-preserving, shallow feature filtering on the original input and may be regarded as a generalized downsampling layer. All remaining skeleton layers are downsamplers that reduce feature-map dimensions. Each skeleton layer is followed by an abstract size-preserving feature extractor containing at least one feature-extraction layer; these modules constitute the specialized feature-extraction process. The final feature extractor is followed by the fully connected and classification layers. Convolutional kernels are denoted by $*_{s,p}^{c,k}$, where $c$ is the number of output channels, $k$ the kernel size, $s$ the stride, and $p$ the padding.

\subsection{The Abstract DenseMedic Architecture: OreoDown}

Building on receptive-field expansion through increased stride, this study first abstracts the DenseMedic architecture into the OreoDown framework. ``OreoDown'' denotes the sandwich-shaped, stepwise downsampling framework composed of skeleton layers and feature-extraction modules. It provides a basis for designing large-receptive-field downsampling FCNs and can be adapted to different segmentation tasks by instantiating appropriate skeleton layers and feature extractors.

\begin{figure}[tbp]
\centering
\includegraphics[width=\textwidth]{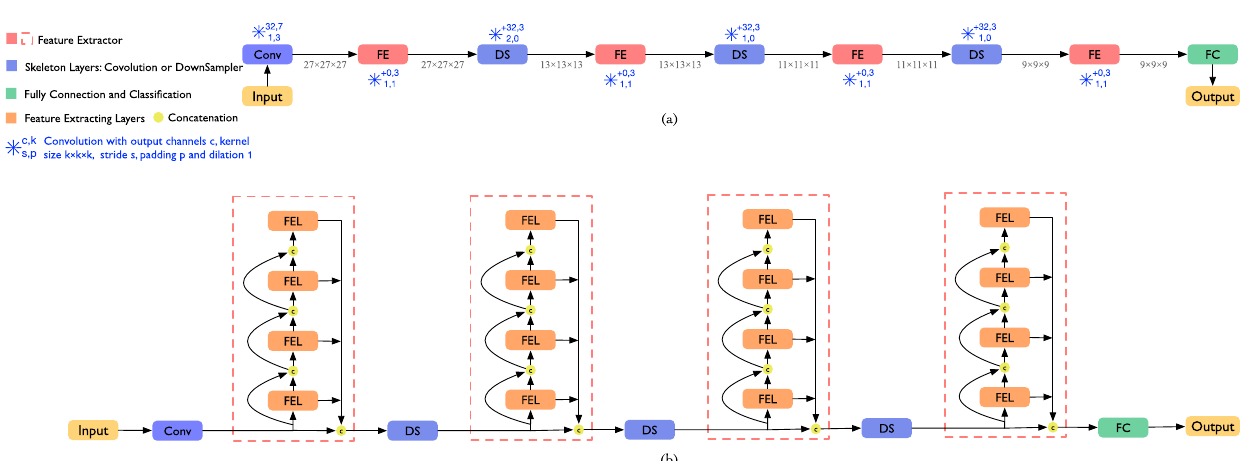}
\caption{Network architectures with an inheritance relationship: (a) the OreoDown downsampling method and (b) DenseMedic.}
\label{fig:oreodown-densemedic}
\end{figure}

In Equation~\eqref{eq:receptive-field}, the cumulative coefficient $\prod_{i=1}^{l-1}\boldsymbol{\tau}_{i}$ is the product of the strides of the first $l-1$ convolutional layers. Receptive-field growth at a layer is therefore affected by the strides of all preceding layers. Increasing the stride $\boldsymbol{\tau}_i$ of several early convolutions, where $i$ is small, multiplies the cumulative coefficient and accelerates receptive-field growth in all later convolutions. For fixed input and output dimensions, however, a large-stride convolution causes a substantial decrease in feature-map dimensions. Downsampling from input to output dimensions can then be completed along a very short path, such as the DS layers in Figure~\ref{fig:oreodown-densemedic}a. The resulting loss of depth would offset the receptive-field benefit of the increased growth rate.

To preserve network depth and convert the increased growth rate into an effective receptive-field increase, OreoDown applies larger convolutional strides in shallow layers and inserts a size-preserving feature extractor after every downsampling operation, producing a stepwise descent. Through their learned convolutional weights, the feature-extraction layers can adjust the scale of local image information required by structures of different sizes within a broader global context. This preserves precise segmentation of small structures even with a large receptive field. This subsection first instantiates the OreoDown skeleton and uses the simplest possible feature extractors to test the effectiveness of this large-scale global and precise local sandwich design. The next subsection introduces deeper feature extractors and thereby instantiates OreoDown as DenseMedic.

The size-changing convolutions of the OreoDown skeleton are instantiated as the purple layers in Figure~\ref{fig:oreodown-densemedic}a. Feature-map dimensions are first reduced by a large-stride convolution, which produces a multiplicative reduction and accelerates receptive-field growth at an early stage. Subsequent unpadded convolutions with a smaller stride reduce the dimensions approximately linearly to the output size. Before the first large-stride convolution, a size-preserving layer with a relatively large kernel is added. This layer is regarded as a generalized downsampler and forms part of the OreoDown skeleton together with the size-changing downsamplers. By filtering the original input with a large kernel, it obtains sufficient low-dimensional features at the original scale and reduces information loss from the rapid decrease in feature-map size. It is therefore an essential safeguard. Because an input MR image has few channels, moderately increasing the kernel size of this layer has little effect on the total parameter count.

For a T1-MR input patch of size $v_0^3$, the generalized downsampling layer first applies a kernel of size $k_0^3$ while preserving the feature-map size $v_0^3$. Several downsamplers then use kernels of size $k_1^3$. The first downsampler has stride $2t$ and zero padding and reduces the feature-map dimensions in one step to $v_1^3$, where
\begin{equation}
v_1=\left\lfloor\frac{v_0}{2t}\right\rfloor.
\end{equation}
Subsequent downsamplers use stride $t$ and zero padding, reducing the dimensions progressively to $v_2^3$ and finally $v_3^3$, where
\begin{equation}
v_2=\frac{v_1-k_1+1}{t},\qquad
v_3=\frac{v_2-k_1+1}{t},
\end{equation}
and $t$ is the basic stride coefficient.

The size-preserving feature-extraction process is first instantiated in its simplest form, shown by the red modules in Figure~\ref{fig:oreodown-densemedic}a. A single size-preserving convolution is placed after every skeleton layer. Its kernel size is $k_1^3$, the same as that of each downsampler; its stride is 1 and its padding is $(k_1-1)/2$. Every convolution is followed by a parametric rectified linear unit (PReLU)\cite{ref46} to introduce a nonlinear mapping. This is the shallowest and simplest network within the OreoDown framework and therefore represents a lower bound on its segmentation performance. Evaluating this lower-bound network tests whether increasing stride within OreoDown effectively expands the receptive field. DenseMedic inherits and extends this architecture.

\subsection{From OreoDown to DenseMedic}

The feature extractors can be redesigned on top of the instantiated skeleton to introduce additional feature-extraction layers. Greater depth supplies global information over a larger receptive field, while richer feature extractors provide more precise local information, potentially improving segmentation further. In image classification, DenseNet propagates the feature distribution of the current layer forward to later feature spaces, facilitating greater depth, encouraging feature reuse, and mitigating vanishing gradients. OreoDown, derived from DeepMedic, is lightweight enough to permit a larger minibatch under the same GPU-memory constraint, allowing the batch-normalization (BN) layers of DenseNet\cite{ref47} to constrain the solution space more effectively. Inspired by the classification setting\cite{ref18}, this study instantiates each OreoDown feature extractor as a DenseBlock\cite{ref40} composed of densely connected feature-extraction layers (FELs), yielding the more accurate DenseMedic architecture. The multiple FELs in each red box of Figure~\ref{fig:oreodown-densemedic}b extend the single red feature extractor of Figure~\ref{fig:oreodown-densemedic}a.

\begin{figure}[tbp]
\centering
\includegraphics[width=0.80\textwidth]{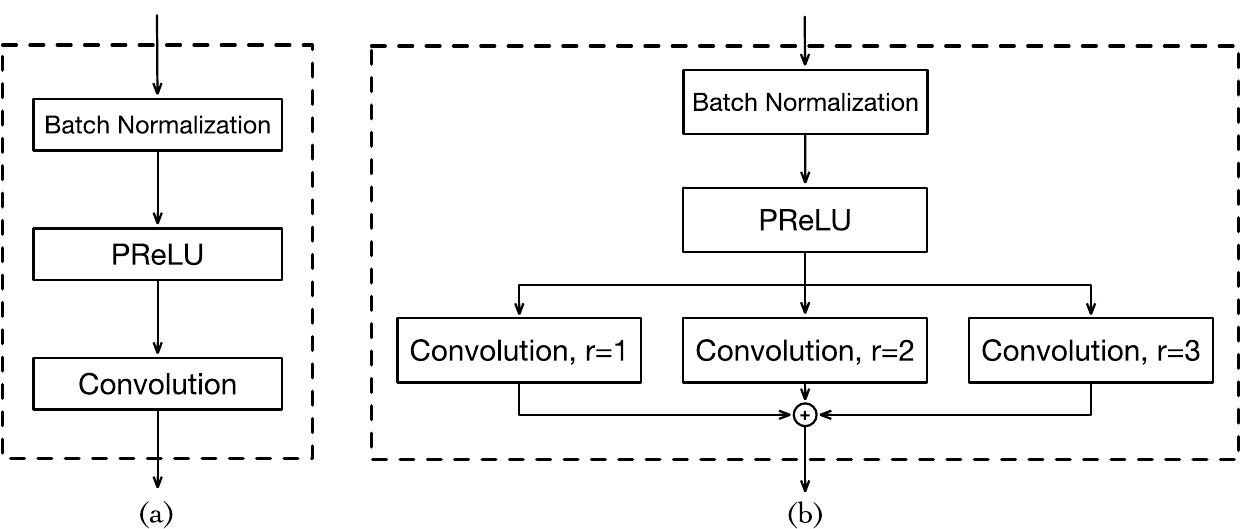}
\caption{DenseMedic layer structures: (a) the original layer and (b) the hybrid dilated-convolution layer.}
\label{fig:densemedic-layers}
\end{figure}

Suppose that DenseMedic groups $m$ densely connected FELs into one module. If layer $n$ is a downsampler with output $x_n$, the output of layer $l$ in the immediately following DenseBlock is
\begin{equation}
x_{n+l}=f_{n+l}^{1}\left([x_n,x_{n+1},\ldots,x_{n+l-1}]\right).
\end{equation}
Here, $f_{n+l}^{r}$ is the mapping at layer $n+l$ and comprises batch normalization, PReLU activation, and a convolution with dilation rate $r$, as shown in Figure~\ref{fig:densemedic-layers}a. Because each module contains $m$ FELs, its output, which is also the input to downsampling layer $n+m+1$, is
\begin{equation}
I_{n+m+1}=[x_n,x_{n+1},\ldots,x_{n+m}].
\end{equation}
The brackets denote channel-wise concatenation. The output $x_{n+m+1}=f_{n+m+1}^{1}(I_{n+m+1})$ becomes the initial input to the next DenseBlock. Batch normalization is included in the mapping of every DenseMedic layer to accelerate convergence and prevent overfitting. With the exception of the first generalized downsampler and the final fully connected and classification layer, which omits BN, all layers use the structure $f_n^r$ illustrated in Figure~\ref{fig:densemedic-layers}a.

\subsection{Hybrid Dilated Convolution in DenseMedic}

The DenseMedic mapping $f_n^r$ is further modified using hybrid dilated convolution to enlarge the receptive field without a large increase in parameters. In a hybrid dilated-convolution layer, the input feature map is first batch-normalized and activated. A parallel group of convolutional kernels with dilation rates $[r_1,\ldots,r_i,\ldots,r_c]$ then filters the result, and their output feature maps are added to form the output of the layer. The output of modified DenseMedic layer $n$ is therefore
\begin{equation}
x'_n=F_n(I_n)=f_n^{r_1}(I_n)+f_n^{r_2}(I_n)+\cdots+f_n^{r_c}(I_n).
\end{equation}
Apart from dilation rate and the corresponding padding, all parallel kernels have identical size, output-channel count, stride, and other parameters. The parallel mappings $f_n^{r_1},f_n^{r_2},\ldots,f_n^{r_c}$ share the same BN and activation layers, as shown in Figure~\ref{fig:densemedic-layers}b.

To prevent missed features, the dilation-rate combination $[r_1,\ldots,r_i,\ldots,r_c]$ must satisfy two conditions: (1) the rates must have no common divisor, and (2), for
\begin{equation}
M_i=\max(r_{i+1}-2r_i,\,2r_i-r_{i+1},\,r_i),
\end{equation}
$M_2\leq k$, where $k$ is the convolutional-kernel size. Interlayer connectivity and input--output relationships remain unchanged from the preceding subsection. With the exception of the first generalized downsampler, all FELs and DSs in DenseMedic are replaced by the hybrid dilated-convolution layer described here.

\subsection{Parameter Selection and Other Details}

Patch-based segmentation was used to address the GPU-memory demands of three-dimensional images and convolutions. A smaller difference between the input and output patch sizes of a downsampling FCN reduces the number of segmentation iterations and shortens inference time, whereas, within a certain range, a larger difference can improve accuracy. In the original DeepMedic study\cite{ref17}, Kamnitsas et al. found that increasing the input T1-MR patch size $v_0$ improved accuracy when $v_0\leq25$, whereas further enlargement beyond 25 yielded no additional improvement. To balance speed and accuracy and permit a controlled comparison between DenseMedic and DeepMedic, this study sets $v_0=27$, as in LiviaNET, the application of DeepMedic to subcortical segmentation by Dolz et al.\cite{ref18}.

The DenseMedic parameter combination was selected according to segmentation performance on IBSR and MRBrainS18 while controlling model scale. For the generalized downsamplers instantiated in Section~3.2.1, the first generalized downsampler uses $k_0=7$, the downsamplers use $k_1=3$, and the basic stride coefficient is $t=1$. The resulting output-patch size is $v_3=9$. Parameters of the simplest and shallowest validation network within OreoDown are marked by asterisks in Figure~\ref{fig:oreodown-densemedic}a. In Section~3.2.2, $m=4$, meaning that every FE contains four densely connected FELs. Each FEL uses $*_{1,1}^{16,3}$ and therefore adds 64 channels to the input feature map within an instantiated FE. A downsampler adds half this number, or 32 channels, while retaining the remaining parameters of the downsampling convolutions in Section~3.2.1. In Section~3.2.3, the hybrid dilation-rate combination is $[1,2,3]$ for every FEL and DS. Finally, the fully connected and classification layers are the same as those of LiviaNET. The fully connected layers use convolutional specifications $*_{1,0}^{400,1}$, $*_{1,0}^{200,1}$, and $*_{1,0}^{160,1}$.

The optimal parameter combination may differ between datasets, in which case the DenseMedic parameters should be adjusted to the data. Adjustable choices include the number of FELs in each DenseBlock, the output-channel counts of the FELs and downsamplers, and the dilation-rate combination in each hybrid dilated-convolution layer. Because the increased-stride design expands the receptive field, other input and output sizes and their corresponding downsampler parameters can also be explored within the OreoDown framework.

\section{Experimental Results}

\subsection{Training, Validation, and Testing}

Adaptive moment estimation (Adam)\cite{ref48} was used for optimization because it adaptively constrains the learning rate and accelerates parameter convergence. The final output layer uses the softmax activation
\begin{equation}
P(S_i)=\frac{e^{g_i}}{\sum_k e^{g_k}},
\end{equation}
where $i$ denotes one of the $k$ classes, $k\in[0,C)$, $C$ is the number of classes, and $g_i$ is the network output for class $i$. The corresponding cross-entropy loss is
\begin{equation}
\mathrm{CE}(p,q)=-\sum_{i=1}^{C}p_i\log(q_i),
\end{equation}
where $p_i$ is the ground-truth value and $q_i$ the predicted probability for class $i$. The initial learning rate $LR$ was 0.001. At epoch $epo$, the learning rate $lr$ was calculated as\cite{ref49}
\begin{equation}
lr=LR\left(1-\frac{epo}{maxepo}\right)^{0.9}.
\end{equation}
The maximum number of epochs, $maxepo$, was 50. Because BN is sensitive to minibatch variation, a minibatch size of 24 was used in every experiment.

\subsection{Results and Analysis}

DenseMedic was evaluated on IBSR and MRBrainS18 and compared with methods reported in related studies. Segmentation was assessed using DSC, IoU, HSD95, and ASD. Better segmentation corresponds to larger DSC and IoU values and smaller HSD95 and ASD values. Because the ``background'' and ``other'' categories do not refer to clearly defined individual structures and were already segmented with high accuracy, their measurements do not clearly quantify network performance. These two classes were therefore excluded from all reported metrics.

Table~\ref{tab:densemedic-ablation} presents ablation studies of the three proposed components on both datasets. For every metric on each dataset, each successive architectural change on the left produced an improvement in the measurements on the right. Relative to the work of Dolz et al.\cite{ref18}, this study increased the per-layer channel increment of DeepMedic from 25 to 32. This both improved accuracy and aligned the basic parameters with the subsequent networks; the modified DeepMedic was therefore used as the baseline. Relative to this baseline, the final DenseMedic increased DSC and IoU by 2.79\% and 4.25\%, respectively, on IBSR, and by 1.30\% and 2.14\% on MRBrainS18, indicating greater spatial overlap between predicted and ground-truth structures. HSD95 and ASD were reduced by approximately sevenfold and twofold on IBSR and fivefold and twofold on MRBrainS18, indicating more accurate delineation of structural contours. The standard deviations across images also decreased, indicating greater robustness.

\begin{table}[!htbp]
\centering
\caption{Ablation study of DenseMedic on (a) IBSR and (b) MRBrainS18. Values are mean $\pm$ standard deviation.}
\label{tab:densemedic-ablation}
\scriptsize
\setlength{\tabcolsep}{2.5pt}
\begin{tabular}{@{}ccccrrrr@{}}
\toprule
Baseline & New DS & Dense FE & HDC & DSC & IoU & HSD95 & ASD \\
\midrule
\multicolumn{8}{c}{\textit{(a) IBSR}} \\
$\checkmark$ & & & & $0.8639\pm0.0032$ & $0.7648\pm0.0045$ & $14.1656\pm2.5686$ & $2.0194\pm0.1950$ \\
$\checkmark$ & $\checkmark$ & & & $0.8719\pm0.0026$ & $0.7766\pm0.0039$ & $7.7257\pm1.7783$ & $1.5487\pm0.1189$ \\
$\checkmark$ & $\checkmark$ & $\checkmark$ & & $0.8829\pm0.0025$ & $0.7934\pm0.0037$ & $7.4359\pm2.7295$ & $1.2984\pm0.2159$ \\
$\checkmark$ & $\checkmark$ & $\checkmark$ & $\checkmark$ & $\mathbf{0.8918\pm0.0024}$ & $\mathbf{0.8073\pm0.0037}$ & $\mathbf{1.9824\pm0.1791}$ & $\mathbf{0.8820\pm0.0518}$ \\
\addlinespace
\multicolumn{8}{c}{\textit{(b) MRBrainS18}} \\
$\checkmark$ & & & & $0.8736\pm0.0036$ & $0.7770\pm0.0056$ & $6.5857\pm1.5270$ & $1.3016\pm0.1202$ \\
$\checkmark$ & $\checkmark$ & & & $0.8831\pm0.0021$ & $0.7928\pm0.0034$ & $2.2793\pm0.0977$ & $0.7158\pm0.0080$ \\
$\checkmark$ & $\checkmark$ & $\checkmark$ & & $0.8855\pm0.0031$ & $0.7967\pm0.0052$ & $1.3274\pm0.0856$ & $0.6192\pm0.0327$ \\
$\checkmark$ & $\checkmark$ & $\checkmark$ & $\checkmark$ & $\mathbf{0.8866\pm0.0018}$ & $\mathbf{0.7984\pm0.0030}$ & $\mathbf{1.2492\pm0.0410}$ & $\mathbf{0.5706\pm0.0189}$ \\
\bottomrule
\end{tabular}
\end{table}

\begin{figure}[!htbp]
\centering
\includegraphics[width=\textwidth]{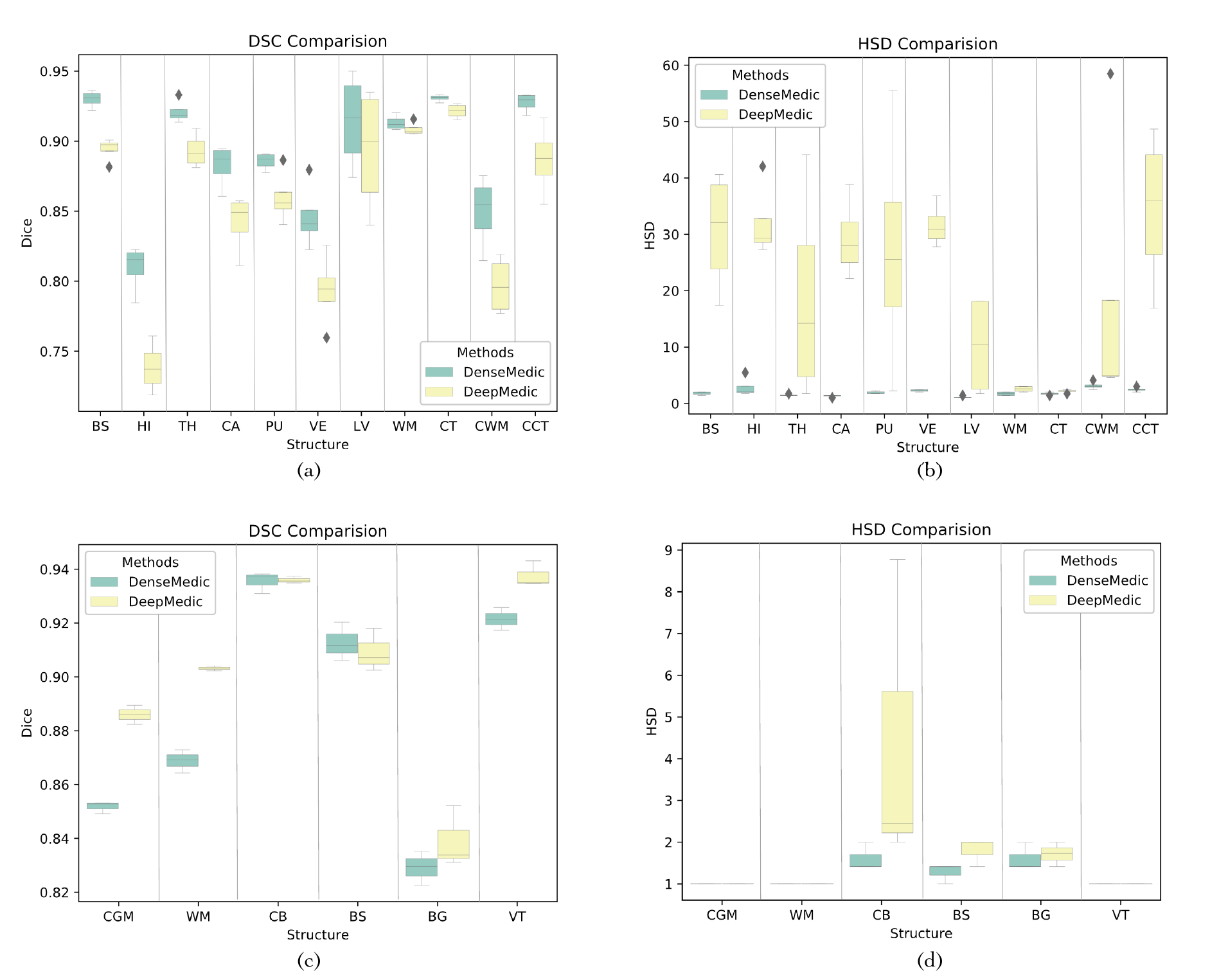}
\caption{Segmentation results of DenseMedic and DeepMedic for individual subcortical structures. IBSR abbreviations: brain stem (BS), hippocampus (HI), thalamus (TH), caudate (CA), putamen (PU), ventral diencephalon (VE), lateral ventricle (LV), cerebral white matter (WM), cerebral cortex (CT), cerebellar white matter (CWM), and cerebellar cortex (CCT). MRBrainS18 abbreviations: cortical gray matter (CGM), white matter (WM), cerebellum (CB), brain stem (BS), basal ganglia (BG), and ventricle (VT). (a) DSC on IBSR; (b) HSD95 on IBSR; (c) DSC on MRBrainS18; and (d) HSD95 on MRBrainS18.}
\label{fig:densemedic-structures}
\end{figure}

Table~\ref{tab:densemedic-comparison} lists the four metrics obtained with leading FCN algorithms on IBSR and MRBrainS18. Every comparator was reimplemented according to its published architecture. On IBSR, the reproduced DeepMedic framework segmented more structures than the original study of Dolz et al.\cite{ref18}; for the three shared structures of thalamus, caudate, and putamen, its DSC values were also higher than the originally reported 0.83, 0.69, and 0.74, as shown in Figure~\ref{fig:densemedic-structures}a. To improve the controlled comparison, BN was inserted before the activation functions of the two U-shaped networks, UNet and VoxResNet. To reduce the influence of differing patch sizes, both U-shaped networks used $32\times32\times32$ patches, close to DenseMedic's $27\times27\times27$ patches. Their training- and validation-patch centers were identical to those of DenseMedic.

\begin{table}[!htbp]
\centering
\caption{DenseMedic comparisons on the single-modal T1 data of IBSR and MRBrainS18. Values are mean $\pm$ standard deviation.}
\label{tab:densemedic-comparison}
\scriptsize
\setlength{\tabcolsep}{2.6pt}
\resizebox{\textwidth}{!}{%
\begin{tabular}{@{}lrrrrrrrr@{}}
\toprule
& \multicolumn{4}{c}{IBSR} & \multicolumn{4}{c}{MRBrainS18} \\
\cmidrule(lr){2-5}\cmidrule(l){6-9}
Method & DSC & IoU & HSD95 & ASD & DSC & IoU & HSD95 & ASD \\
\midrule
UNet\cite{ref12} & $0.8790\pm0.0038$ & $0.7889\pm0.0055$ & $2.5571\pm0.5537$ & $0.9488\pm0.0922$ & $0.8759\pm0.0105$ & $0.7820\pm0.0173$ & $2.0153\pm0.4307$ & $0.7561\pm0.0592$ \\
VoxResNet\cite{ref13} & $0.8772\pm0.0023$ & $0.7861\pm0.0038$ & $21.4410\pm3.0463$ & $3.0210\pm0.1323$ & $0.8743\pm0.0162$ & $0.7798\pm0.0260$ & $2.0476\pm0.4190$ & $0.7418\pm0.1512$ \\
DeepMedic\cite{ref18} & $0.8571\pm0.0027$ & $0.7544\pm0.0040$ & $21.5226\pm6.7981$ & $2.7501\pm0.4351$ & $\mathbf{0.9018\pm0.0008}$ & $\mathbf{0.8229\pm0.0010}$ & $1.8214\pm0.5022$ & $0.6145\pm0.0324$ \\
HyperDense-Net\cite{ref19} & $0.8233\pm0.0100$ & $0.7084\pm0.0132$ & $23.4084\pm1.1767$ & $4.3981\pm0.4418$ & $0.8987\pm0.0036$ & $0.8176\pm0.0059$ & $1.5064\pm0.0900$ & $0.6042\pm0.0336$ \\
DenseMedic & $\mathbf{0.8918\pm0.0024}$ & $\mathbf{0.8073\pm0.0037}$ & $\mathbf{1.9824\pm0.1791}$ & $\mathbf{0.8820\pm0.0518}$ & $0.8866\pm0.0018$ & $0.7984\pm0.0030$ & $\mathbf{1.2492\pm0.0410}$ & $\mathbf{0.5706\pm0.0189}$ \\
\bottomrule
\end{tabular}}
\end{table}

As shown in Table~\ref{tab:densemedic-comparison}, DenseMedic improved all four metrics relative to the comparators. Compared with UNet on IBSR, its DSC reached 0.8918, an improvement of 1.28\%; IoU reached 0.8073, an improvement of 1.84\%; HSD95 decreased by approximately 0.5747 to 1.9824; and ASD decreased by approximately 0.0668 to 0.8820.

To examine the gains in more detail, DeepMedic and DenseMedic were compared quantitatively for the 11 IBSR structures in Figure~\ref{fig:densemedic-structures}a and~b. DenseMedic produced substantial DSC improvements for every structure. It also greatly reduced HSD95 and its variance, indicating that precise segmentation was more stable across images. For the six MRBrainS18 structures shown in Figure~\ref{fig:densemedic-structures}c and~d, DenseMedic produced particularly large gains for the cerebellum and brain stem, with higher DSC and lower HSD95. HSD95 decreased for every other structure as well, indicating more accurate contours, although DSC decreased slightly for the other structures.

\begin{figure}[!htbp]
\centering
\includegraphics[width=\textwidth]{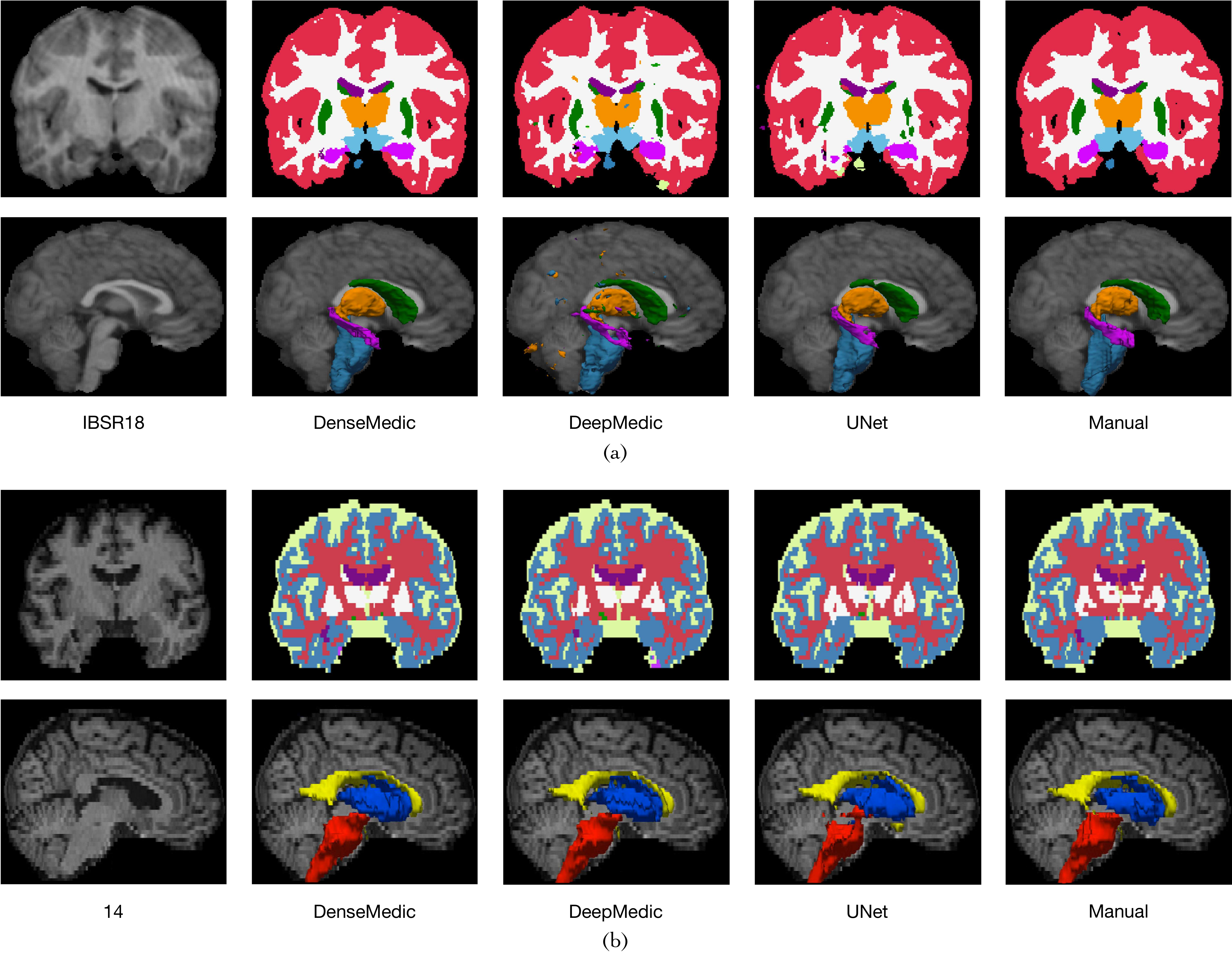}
\caption{Segmentation results of different networks in two orientations: (a) visual comparison on IBSR and (b) visual comparison on MRBrainS18.}
\label{fig:densemedic-visual}
\end{figure}

Figure~\ref{fig:densemedic-visual}a qualitatively compares three-dimensional segmentations on IBSR in two orientations. The first row shows planar views of all segmented structures in the coronal plane. DenseMedic produces fewer bubble-like artifacts in the red cerebral cortex, fewer extraneous colors in the white cerebral matter, and more regular contours for the smaller structures. The second row shows three-dimensional forms of the brain stem, hippocampus, thalamus, and caudate in the sagittal orientation. DenseMedic fills cavities in the hippocampus and thalamus and missing regions in the caudate without introducing additional fragments.

Figure~\ref{fig:densemedic-visual}b provides a comparable qualitative assessment on MRBrainS18. In the coronal views of the first row, DenseMedic more accurately segments the purple ventricle and green brain-stem pixels in the lower-left portion of the image. The sagittal views in the second row show the basal ganglia, ventricle, and brain stem. Compared with DeepMedic, DenseMedic avoids excessive surface folds in the yellow ventricle and blue basal ganglia, consistent with its lower HSD95. Compared with UNet, it avoids the defect at the superior end of the red brain stem and therefore segments both the region and its contour more accurately.

\section{Discussion}

Among FCN-based networks for subcortical segmentation, DenseMedic predicts spatial probability distributions directly from small feature maps. It therefore reduces GPU-memory demand during training and inference while retaining high accuracy, combining accuracy with a lightweight architecture and achieving advanced performance on IBSR and MRBrainS18. The ablation results in Table~\ref{tab:densemedic-ablation} show clear and consistent improvements in DSC, IoU, HSD95, and ASD at every stage on both datasets. On IBSR T1-MR images, Table~\ref{tab:densemedic-comparison} shows that DenseMedic considerably outperforms both DeepMedic and HyperDense-Net, which densely connects all DeepMedic layers, and also outperforms the classic UNet and VoxResNet. On MRBrainS18 T1-MR images, DenseMedic ranks third in DSC behind DeepMedic and HyperDense-Net, but its measurements remain sufficient to preserve regional accuracy in the visual three-dimensional results. It outperforms all four comparators in the contour-oriented HSD95 and ASD metrics.

The per-structure DSC comparisons in Figure~\ref{fig:densemedic-structures} show greater overlap with the ground truth for all 11 IBSR structures and two MRBrainS18 structures. The HSD95 comparisons show improved contour accuracy for all 11 IBSR structures and all six MRBrainS18 structures. DenseMedic therefore identifies the regional locations of tissues more accurately than the original networks and produces more plausible contours. In Figure~\ref{fig:densemedic-visual}a, the small receptive field of DeepMedic leads it to misclassify some background voxels in locations occupied by large structures such as cerebral gray and white matter as small structures such as the thalamus and brain stem. The resulting fragments impair contour delineation and are consistent with its high HSD95 and ASD. UNet produces more gaps and cavities in the three-dimensional caudate, thalamus, and hippocampus, suggesting that its large-scale output features do not adapt well to small subcortical structures. Neither defect is apparent in the DenseMedic results, consistent with its strong regional and contour performance.

The visual comparison in Figure~\ref{fig:densemedic-visual}b further indicates that precise local perception within broad global information enables DenseMedic to address both large and small structures during whole-brain segmentation. It reduces both missing regions, as seen in the basal ganglia and brain stem of DeepMedic and the brain stem of UNet, and fragments, as seen in the ventricles of DeepMedic and the ventricle, basal ganglia, and brain stem of UNet. Although DenseMedic has lower DSC than DeepMedic for four structures in Figure~\ref{fig:densemedic-structures}c, the three-dimensional comparisons for two of these structures, the ventricle and basal ganglia, appear closer to the corresponding ground truth. HSD95 was therefore considered more informative for these four structures. Consistent with the MRBrainS18 results in Table~\ref{tab:densemedic-comparison}, DenseMedic retained advanced segmentation performance in HSD95 and ASD compared with the other four methods.

Equation~\eqref{eq:receptive-field} provides a basis for discussing receptive field. In DeepMedic\cite{ref18}, the input and output patch sizes are $27^3$ and $9^3$, respectively, with $\boldsymbol{\kappa}_i=[3,3,3]$, $\boldsymbol{\tau}_i=[1,1,1]$, and $i=1,2,\ldots,9$. The receptive field of each output voxel in the original input is therefore
\begin{equation}
\boldsymbol{\varphi}_9=[1,1,1]+9\left([3,3,3]-[1,1,1]\right)\circ[1,1,1]=[19,19,19].
\end{equation}
Thus, each voxel in the $9^3$ output patch uses only a $19^3$ region of the $27^3$ input patch, or 34.85\% of the input voxels. Image information available for each prediction is therefore reduced relative to the original input. The loss is more consequential when the physical voxel dimensions are considered. For example, IBSR\cite{ref18} has voxel dimensions of $0.8371\times0.8371\times1.5\,\mathrm{mm}^3$. Losing eight voxels along the $z$ axis means that information from $0.15\times8=1.2\,\mathrm{cm}$ of tissue thickness is ignored during prediction. Tissue of this thickness can affect clinical assessment of a brain structure or lesion, motivating the OreoDown design based on increased stride.

In the shallowest OreoDown instantiation marked by asterisks in Figure~\ref{fig:oreodown-densemedic}a, the six layers after the third, large-stride convolution contribute a receptive-field increment of
\begin{equation}
\boldsymbol{\varphi}'_8-\boldsymbol{\varphi}'_2
=6\left([3,3,3]-[1,1,1]\right)\circ[2,2,2]=[24,24,24].
\end{equation}
This is one third larger than the increment across nine layers of the original DeepMedic,
\begin{equation}
\boldsymbol{\varphi}_9-\boldsymbol{\varphi}_0
=9\left([3,3,3]-[1,1,1]\right)\circ[1,1,1]=[18,18,18].
\end{equation}
Including the first two layers, the lower bound of the OreoDown receptive field is
\begin{align}
\boldsymbol{\varphi}'_8
&=[1,1,1]+\left([7,7,7]-[1,1,1]\right)\circ[1,1,1] \notag\\
&\quad+\left([3,3,3]-[1,1,1]\right)\circ[1,1,1]+[24,24,24] \notag\\
&=[33,33,33].
\end{align}
The receptive field of an output voxel therefore essentially covers the full input patch. DenseMedic enlarges it further. In deep networks, padding can cause the theoretical receptive field to exceed the input resolution. The fully connected layers of this FCN do not affect the receptive field because their kernels have size 1.

One limitation of DenseMedic is that the large-stride rapid downsampling occurs too early, causing an abrupt change from $v_0^3$ to $v_1^3$. Although the original-scale feature extractor captures as much low-dimensional information as possible, the rapid dimensional change still loses semantic information at intermediate scales. The input and output dimensions of the rapid downsampler have a multiplicative relationship, $v_1=\lfloor v_0/(2t)\rfloor$. Introducing an appropriate amount of linear downsampling before rapid downsampling permits a new DenseMedic input size $v>v_0$, such that
\begin{equation}
\frac{v_0-v_1}{v-v_3}<\frac{v_0-v_1}{v_0-v_3}.
\end{equation}
This reduces the proportion of the overall network size change attributable to the abrupt span $v_0-v_1$, weakening the effect of the multiplicative reduction. It also deepens DenseMedic and may improve performance within OreoDown. This issue can therefore be addressed by a fine-adjustment--rapid--fine-adjustment downsampling sequence that delays the large-stride operation appropriately. Future work should also seek improved feature-extractor instantiations that reduce the parameter count while improving segmentation.

\section{Chapter Summary}

To improve both regional and contour accuracy in subcortical segmentation of T1-MR images, this chapter proposed DenseMedic. A new rapid-downsampling step separates downsampling from feature extraction; densely connected FELs strengthen perception of contextual semantic information; and parallel hybrid dilated convolutions enlarge the receptive field while refining the coarse feature sampling of dilated kernels. Experiments on IBSR and MRBrainS18 showed that each modification improved the segmentation of both regions and contours. Compared with results from previous methods, DenseMedic performed strongly on all four conventional metrics. Its subcortical regions were more accurate, its contours more regular, and both findings were supported by visual assessment.

\chapter{Subcortical Brain-Structure Segmentation Using an Alternate Connected Neural Network}

\section{Introduction}

Chapter~3 introduced OreoDown, a design framework for downsampling FCNs. Rapid downsampling separates the size-changing downsampling process from size-preserving feature extraction, allowing stronger segmentation networks to be designed by choosing how the feature-extraction process is instantiated. DenseMedic subsequently instantiated the predefined downsampling process with densely connected feature extractors and achieved strong segmentation performance. Although each densely connected layer has few output channels, however, its number of input channels increases linearly with network depth. Beyond a certain depth, this can produce explosive growth in parameter count\cite{ref40}, increase the risk of overfitting during feature estimation, and impair segmentation. A standardized interlayer connection pattern that substantially reduces the parameter count while improving performance would therefore provide a better instantiation of OreoDown. This study proposes alternate connections between layers and uses them to instantiate OreoDown as an alternate connected neural network (ACNN) with fewer parameters and stronger performance.

As discussed in Section~1.2.2.5, neither DeepMedic nor HyperDense-Net simultaneously achieved state-of-the-art performance in both single- and multimodal segmentation\cite{ref18,ref19}. With the proposed alternate connections, single-path ACNN achieved advanced subcortical segmentation on single-modal T1-MR images using relatively few parameters. The single path was then divided at its center to obtain a dual-path ACNN. This division neither adds nor removes parameters, yet the resulting model achieved advanced performance on the single-modal T1 and dual-modal T1+FLAIR and T1+T1-IR tasks. In multimodal segmentation, the modalities naturally have different input distributions, and dual- or multipath networks, including HyperDense-Net\cite{ref19}, MMAN\cite{ref51}, and ACNN, can assign each modality to a separate path. Given strong single-modal performance, prior information can likewise be introduced so that versions of the same single-modal image with slightly different intensity distributions enter the two ACNN paths, potentially improving single-modal segmentation further\cite{ref52}. The proposed dual-path ACNN is therefore applicable to both single- and multimodal segmentation. Prior-information enhancement is added for single-modal tasks, enabling advanced performance across both settings under one architecture.

\section{Design of ACNN}

Within OreoDown, the abbreviated ACNN architecture is shown in Figure~\ref{fig:acnn-architecture}a. As before, the network consists of purple skeleton layers (SLs), red feature extractors (FEs), and green fully connected and classification layers. A feature extractor follows every skeleton layer. The skeleton comprises the initial low-dimensional convolution and all downsamplers.

\subsection{From OreoDown to ACNN}

Building on OreoDown and the downsampling process instantiated in Section~3.2.1, each feature extractor is first instantiated with $m$ serial size-preserving feature-extraction layers (FELs), producing the initial ACNN in Figure~\ref{fig:acnn-architecture}b. Each red box and its FELs represent one instantiated feature extractor. A feature extractor together with its following downsampler, if present, is defined as an alternate unit (AU), indicated by a brown box. After the network has been divided into AUs, adjacent units are connected alternately. The output-channel count of each layer remains unchanged from the initial instantiation, whereas its input-channel count changes according to the layers connected to it.

Under alternate connectivity, the output of a layer is propagated only to every convolution in its own AU and the immediately following AU, and not to nonadjacent units further forward. Conversely, the input to a convolution comprises outputs of preceding layers in its own AU and the immediately preceding AU, but it receives no features from nonadjacent earlier units. For network layer $L_n$ in AU $A_l$, the output is
\begin{equation}
x_n=f_n\left(
\left[
\left\{x_i\mid L_i\in A_{l-1}\cup A_l,\ i<n\right\}
\right]
\right),
\label{eq:alternate-single}
\end{equation}
where $l\geq1$, $A_0$ contains only the first convolution, $f_n$ is the mapping of layer $n$, and brackets denote feature concatenation. Figure~\ref{fig:acnn-architecture}c samples the input sources of four convolutions under this rule. For each convolutional input, these connections coexist; the layers are therefore not isolated as the simplified drawing might suggest but form an interconnected whole through alternate connectivity, instantiating OreoDown as a single-path ACNN. The FC block receives the concatenated outputs of all preceding layers. When feature maps of different dimensions are concatenated, center cropping is used\cite{ref18,ref19}.

Unlike fully dense connectivity, ACNN partitions the layers into AUs and connects only adjacent units; there are no links that skip over an adjacent unit to a more distant one. Alternate connectivity reduces the feature input of every layer while retaining small-channel outputs and improving segmentation. Because ACNN uses alternate rather than DenseMedic's dense connections to instantiate OreoDown, its layer composition remains the same while its input-channel counts are substantially reduced, lowering the total number of parameters. ACNN inherits the DenseMedic parameter choices in Section~3.2.4. The layerwise parameter counts are compared in Table~\ref{tab:parameter-comparison}. Under other parameter choices, increasing depth within OreoDown would cause the DenseMedic parameter count to grow more rapidly, further widening the difference between the two networks.

During feature perception before the FC block, each ACNN layer needs to perceive local information only within a sufficiently broad, defined range, concentrating semantic extraction at the current scale. The semantic information at a scale need not propagate far through the feature-perception stage; transmission to the next unit is sufficient. Global contextual information is used only during fully connected classification, where outputs from intermediate layers at every scale, each focused on local information, are passed to the final segmentation process. This explicit division between local perception during feature extraction and global perception during final classification gives ACNN stronger dense-inference capability while alleviating the parameter growth caused by dense interlayer connections.

\subsection{Dual-Path ACNN}

Single-path ACNN achieves advanced performance in single-modal medical-image segmentation, but a unified framework is also needed for multimodal tasks. Alternate connectivity makes such a design straightforward. A natural multimodal design contains one path per input modality. To preserve single-modal performance, this multipath construction should not introduce a large parameter increase and ideally should leave the total count unchanged. Because alternate connectivity has already reduced the input-channel count of each layer, $P$ ACNN paths can be constructed by holding every layer's input-channel count constant and reducing its output-channel count to $1/P$ of the original value.

For layer $L_n^p$ in AU $A_l^p$ of path $p$, the output is
\begin{equation}
x_n^p=f_n^p\left(
\left[
\left\{x_i^j\mid L_i^j\in A_{l-1}^j\cup A_l^j,\ i<n,\ j\in[0,P),\ i,j\in\mathbb{N}\right\}
\right]
\right).
\label{eq:alternate-multipath}
\end{equation}
Here, $L_i^j$ ranges over all layers before layer $n$ in $A_{l-1}^j\cup A_l^j$ on path $j$, and the brackets denote center-cropped concatenation of every feature map in the set. Because $p$ does not occur inside the expression passed to $f_n^p$, layer $n$ receives the same input on every path. From a cross-path view, corresponding layer-$n$ inputs are identical and their output channels partition those of the original single-path layer. Corresponding layers on different paths can therefore still be regarded as one architectural layer, so multipath ACNN remains an instantiation of OreoDown.

\begin{figure}[!t]
\centering
\includegraphics[height=0.82\textheight,keepaspectratio]{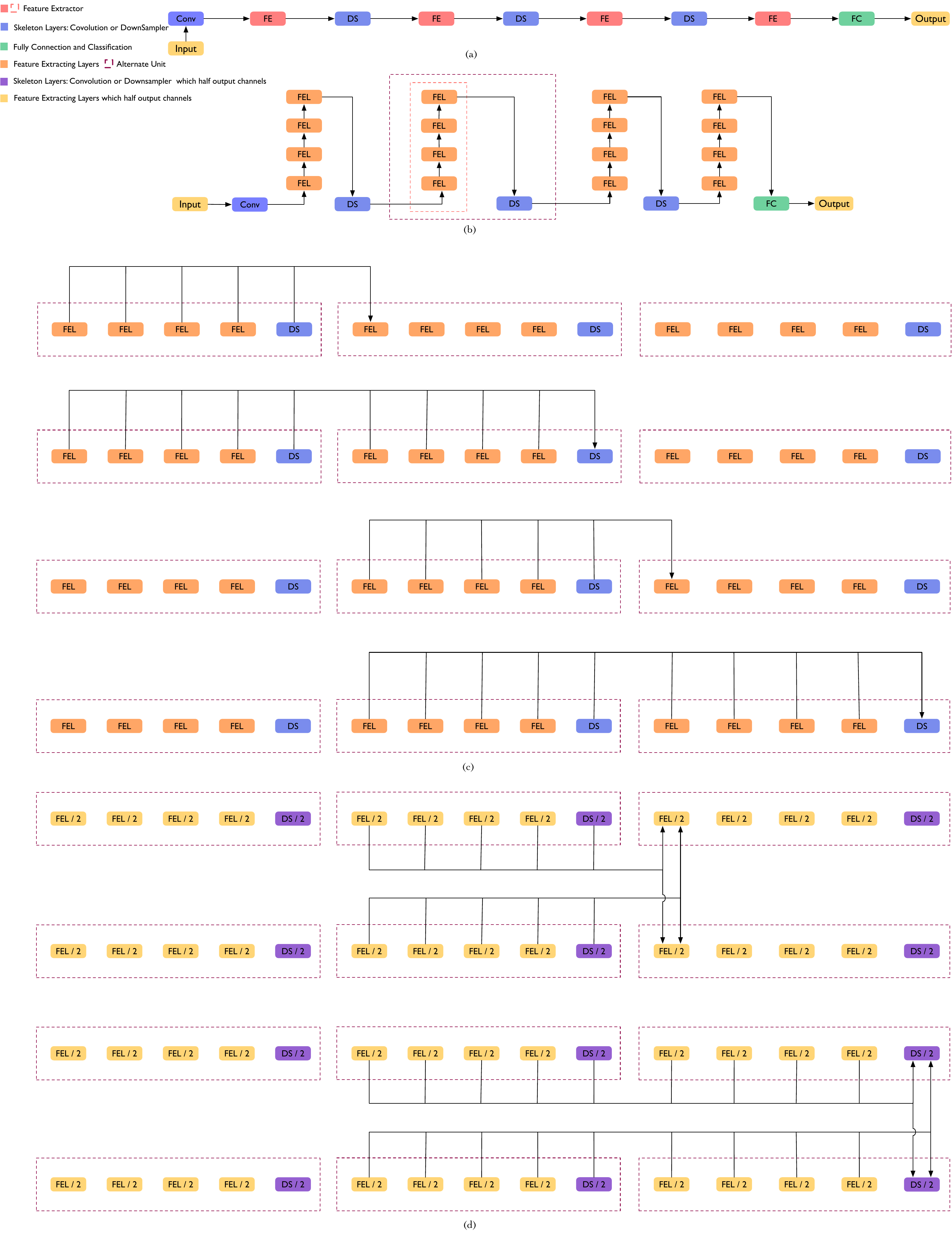}
\caption{Inherited ACNN architectures: (a) the OreoDown framework; (b) the initial ACNN instantiation within OreoDown; (c) alternate connections between layers; and (d) alternate connections between paths and layers in multimodal ACNN ($P=2$).}
\label{fig:acnn-architecture}
\end{figure}

Figure~\ref{fig:acnn-architecture}d extends the final two alternate-connected layers in Figure~\ref{fig:acnn-architecture}c to show cross-path connectivity. Yellow layers are FELs whose input-channel count is unchanged and output-channel count is halved; aubergine layers are similarly modified downsamplers. Center-cropped concatenation joins features both between paths and between layers, visually representing Equation~\eqref{eq:alternate-multipath}. The first layer of the FC block is also divided across paths. Its copy on path $p$ receives global information from all layers of that path. The outputs of this layer from different paths are concatenated and passed to the subsequent unified single-path fully connected and classification layers. The mapping $f_n^p$ consists of group normalization (GN), PReLU activation, and convolution.

Conceptually, multipath ACNN is produced by cutting the single-path ACNN through its center, dividing it into paths, and reconnecting its layers across paths through alternate connections. Different modalities can enter different paths, allowing alternate connectivity to benefit multimodal segmentation. Because path division does not change the parameter count, the same single-modal image can also be supplied to all paths, preserving and improving the single-path advantage. By centrally cutting, dividing, and reconnecting single-path ACNN, the proposed design unifies segmentation problems with different numbers of modalities and produces a multipath network that performs strongly in both single- and multimodal tasks within OreoDown.

\begin{table}[!htbp]
\centering
\caption{Parameter-count comparison between dense and alternate connectivity within OreoDown.}
\label{tab:parameter-comparison}
\scriptsize
\setlength{\tabcolsep}{2.5pt}
\resizebox{\textwidth}{!}{%
\begin{tabular}{@{}llrrr rrr l@{}}
\toprule
& & \multicolumn{3}{c}{DenseMedic} & \multicolumn{3}{c}{ACNN} & \\
\cmidrule(lr){3-5}\cmidrule(lr){6-8}
Layer & Kernel & Input ch. & Output ch. & Parameters & Input ch. & Output ch. & Parameters & Output size \\
\midrule
Conv & $7^3$ & 1 & 64 & 21,952 & 1 & 64 & 21,952 & $27^3$ \\
FEL\_0 & $3^3$ & 64 & 16 & 27,648 & 64 & 16 & 27,648 & $27^3$ \\
FEL\_1 & $3^3$ & 80 & 16 & 34,560 & 80 & 16 & 34,560 & $27^3$ \\
FEL\_2 & $3^3$ & 96 & 16 & 41,472 & 96 & 16 & 41,472 & $27^3$ \\
FEL\_3 & $3^3$ & 112 & 16 & 48,384 & 112 & 16 & 48,384 & $27^3$ \\
DS\_0 & $3^3$ & 128 & 160 & 552,960 & 128 & 96 & 331,776 & $13^3$ \\
FEL\_4 & $3^3$ & 160 & 16 & 69,120 & 160 & 16 & 69,120 & $13^3$ \\
FEL\_5 & $3^3$ & 176 & 16 & 76,032 & 176 & 16 & 76,032 & $13^3$ \\
FEL\_6 & $3^3$ & 192 & 16 & 82,944 & 192 & 16 & 82,944 & $13^3$ \\
FEL\_7 & $3^3$ & 208 & 16 & 89,856 & 208 & 16 & 89,856 & $13^3$ \\
DS\_1 & $3^3$ & 224 & 256 & 1,548,288 & 224 & 96 & 580,608 & $11^3$ \\
FEL\_8 & $3^3$ & 256 & 16 & 110,592 & 160 & 16 & 69,120 & $11^3$ \\
FEL\_9 & $3^3$ & 272 & 16 & 117,504 & 176 & 16 & 76,032 & $11^3$ \\
FEL\_10 & $3^3$ & 288 & 16 & 124,416 & 192 & 16 & 82,944 & $11^3$ \\
FEL\_11 & $3^3$ & 304 & 16 & 131,328 & 208 & 16 & 89,856 & $11^3$ \\
DS\_2 & $3^3$ & 320 & 352 & 3,041,280 & 224 & 96 & 580,608 & $9^3$ \\
FEL\_12 & $3^3$ & 352 & 16 & 152,064 & 160 & 16 & 69,120 & $9^3$ \\
FEL\_13 & $3^3$ & 368 & 16 & 158,976 & 176 & 16 & 76,032 & $9^3$ \\
FEL\_14 & $3^3$ & 384 & 16 & 165,888 & 192 & 16 & 82,944 & $9^3$ \\
FEL\_15 & $3^3$ & 400 & 16 & 172,800 & 208 & 16 & 89,856 & $9^3$ \\
FC\_0 & $1^3$ & 416 & 400 & 166,400 & 608 & 400 & 243,200 & $9^3$ \\
FC\_1 & $1^3$ & 400 & 200 & 80,000 & 400 & 200 & 80,000 & $9^3$ \\
FC\_2 & $1^3$ & 200 & 160 & 32,000 & 200 & 160 & 32,000 & $9^3$ \\
\midrule
\multicolumn{2}{l}{Total parameters} & & & \textbf{7,046,464} & & & \textbf{2,976,064} & \\
\bottomrule
\end{tabular}}
\end{table}

\subsection{Introducing Prior Information for Single-Modal Segmentation}

Multipath ACNN naturally supports multimodal medical-image segmentation and, because path division does not add parameters, retains and improves the single-modal advantage of single-path ACNN. If prior information is introduced so that single-modal T1-MR images containing different spatial semantic information and slightly different distributions are supplied to different ACNN paths, single-modal segmentation may improve further. Similar prior-information integration could also be explored for multimodal tasks in future work.

Wachinger et al.\cite{ref37} noted that a limitation of patch-based deep-learning segmentation is the loss of spatial context, which prevents the model from obtaining useful information for structures with low tissue contrast. They showed that introducing spectral coordinates\cite{ref50} into a conventional patch-wise deep-learning method can supply spatial context and improve segmentation. Such prior information had not been effectively introduced into FCN-based segmentation. With an appropriate design, however, it should also benefit dense-inference networks. Because multipath ACNN requires prior information to vary the spatial distribution of single-modal inputs, this study introduces the spatial context of spectral coordinates into an FCN dense-inference method for the first time, further strengthening ACNN on single-modal T1-MR images.

\subsubsection{Calculation of Spectral Coordinates}

An adjacency matrix $W$ is first constructed from each T1-MR image. Because its weights measure the similarity between every pair of points in the image, $W$ is an extremely large sparse matrix for a three-dimensional medical image and must be stored and processed using a sparse representation. Its dimensions are $(256\times128\times256)^2$ for IBSR and $(240\times240\times48)^2$ for MRBrainS18. The weighting rule was modified so that the image-intensity distribution contributes to the estimated spatial distribution of the spectral coordinates. For any two points $q_i$ and $q_j$, if they are spatial neighbors and both belong to the set of brain-region voxels,
\begin{equation}
W(q_i,q_j)=I(q_i)+I(q_j);
\end{equation}
otherwise, $W(q_i,q_j)=0$. Here, $I$ is the normalized medical image. Point $q_i$ is considered adjacent to $q_j$ when it falls within the $3^3$ neighborhood centered at $q_j$.

\begin{figure}[!htbp]
\centering
\includegraphics[width=\textwidth,keepaspectratio]{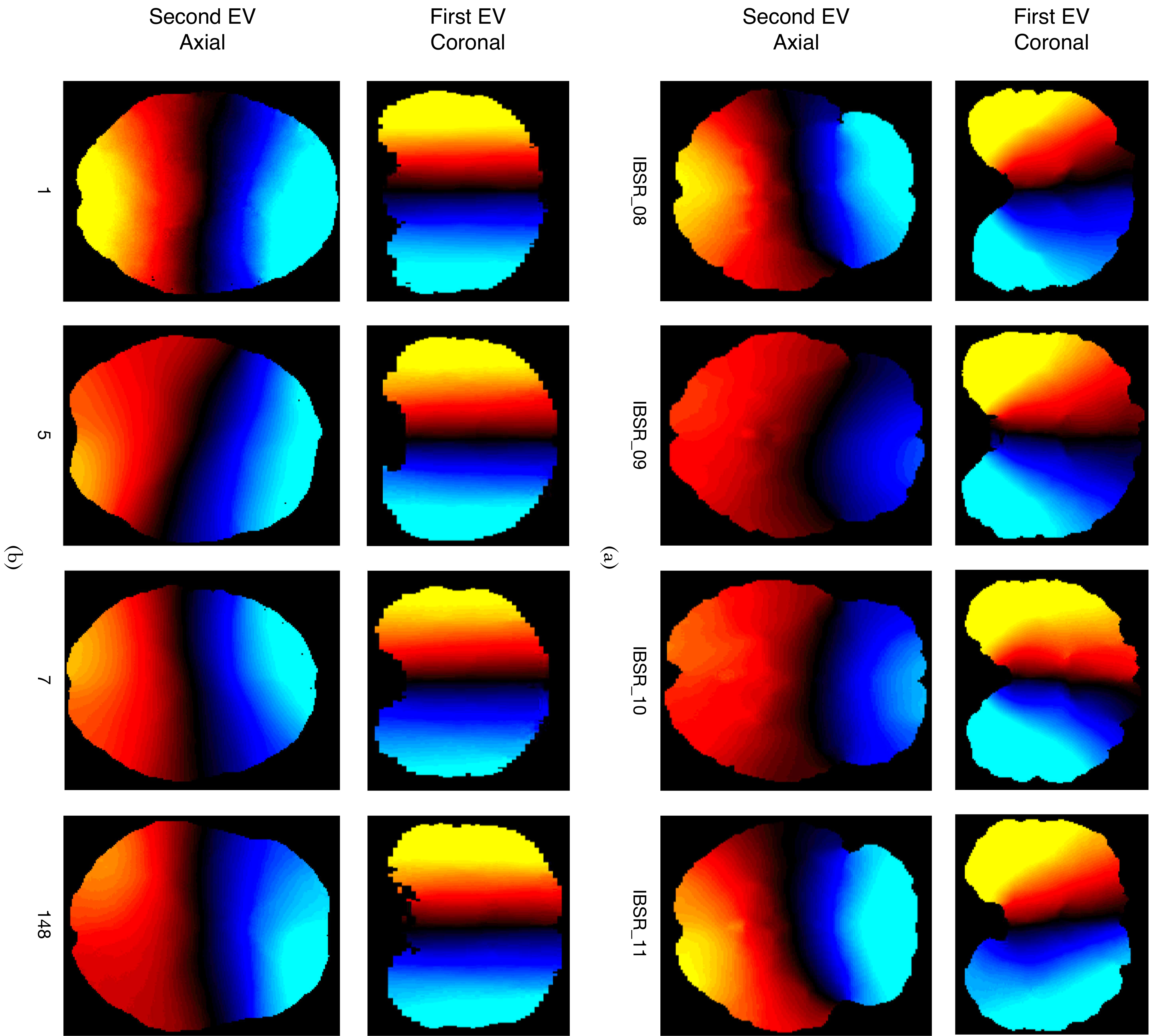}
\caption{Examples of eigenvectors used as spectral coordinates ($P=2$): (a) two spectral coordinates for IBSR and (b) two spectral coordinates for MRBrainS18.}
\label{fig:spectral-coordinates}
\end{figure}

The graph Laplacian is then computed as
\begin{equation}
L=D-W,
\end{equation}
where the degree matrix $D$ is diagonal and $D(i,i)=\sum_j W(i,j)$. Solving the eigenvalue problem
\begin{equation}
Lf=-\lambda f
\end{equation}
gives the first $P$ smallest nonzero eigenvalues $\lambda_0,\lambda_1,\ldots,\lambda_{P-1}$ and their eigenvectors $f_0,f_1,\ldots,f_{P-1}$, where $P$ is the number of ACNN paths. The real component of each eigenvector is reshaped to the dimensions of the original image, producing $P$ spectral-coordinate volumes. These coordinates supply additional spatial semantic information and effectively alter the intensity distribution presented to each path. Figure~\ref{fig:spectral-coordinates} shows two-dimensional views, in the coronal and axial planes, of the two spectral coordinates calculated for four T1-MR images from each of IBSR and MRBrainS18. Their consistent spatial distributions across images support dual-path ACNN segmentation of subcortical tissues in single-modal T1-MR images.

\subsubsection{Spectral Enhancement}

The preceding subsection calculated $P$-dimensional spectral coordinates for the $P$ ACNN paths. These coordinates improve the spatial distribution and introduce semantic information for each path. They are incorporated through spectral enhancement of the T1-MR image. For normalized image $I$ and spectral coordinate $f_i$ on a given path, where $i\in[0,P)\cap\mathbb{N}$,
\begin{equation}
I'=I+\delta
\frac{f_i-\min(f_i)}{\max(f_i)-\min(f_i)}\circ\operatorname{mask}(I),
\label{eq:spectral-enhancement}
\end{equation}
where $\delta$ is the spectral-enhancement coefficient, $\max(f_i)$ and $\min(f_i)$ are the maximum and minimum of eigenvector $f_i$, $\operatorname{mask}(I)$ is the brain mask of image $I$, and $\circ$ is the Hadamard product. Every spectral eigenvector is incorporated into the input of its corresponding path using Equation~\eqref{eq:spectral-enhancement}, thereby spectrally enhancing the MR intensity distribution for that path. Other methods of incorporating the eigenvectors may also be explored to improve single-modal ACNN segmentation further.

\subsection{Parameter Selection and Other Details}

Patch-based segmentation was again used to address the memory requirements of three-dimensional images and convolutions. Within OreoDown, the input and output patch sizes remain as specified in Section~3.2.4: $v_0=27$ and $v_3=9$ after the instantiated downsampling process. Other downsampling parameters follow Sections~3.2.1 and~3.2.4. The ACNN parameter combination was selected according to segmentation performance on IBSR and MRBrainS18. In Section~4.2.1, $m=4$, so each AU contains four FELs and one downsampler, except that the final AU contains only four FELs. Each FEL uses $*_{1,1}^{16,3}$, so the FELs in each AU output 64 channels. When present, the downsampler outputs 1.5 times this value, or 96 channels. Each AU therefore relays 160 feature channels to the next unit.

For the single-modal IBSR task and the single- and dual-modal MRBrainS18 tasks in Section~4.2.2, $P=2$, so a dual-path ACNN is used and every FEL has specification $*_{1,1}^{8,3}$. In the final fully connected classification process, the first fully connected layer on each path uses $*_{1,0}^{320,1}$. After the paths are concatenated, the single-path fully connected layers use $*_{1,0}^{400,1}$, $*_{1,0}^{200,1}$, and $*_{1,0}^{160,1}$, matching the corresponding parts of DeepMedic and DenseMedic. GN uses 16 groups. The spectral-enhancement coefficient in Section~4.2.3 is fixed at $\delta=0.1$. The optimal parameter combination may vary with the dataset and should then be adjusted to its characteristics.

\section{Experimental Results}

\subsection{Training, Validation, and Testing}

The Adam optimizer, final-layer softmax activation, cross-entropy loss, and learning-rate schedule used in the ACNN experiments were the same as those described in Section~3.3.1. Although GN is no longer sensitive to minibatch variation, the minibatch size was fixed at 20 in all experiments.

\subsection{Results and Analysis}

ACNN was evaluated on IBSR and MRBrainS18 and compared with results from related methods. Segmentation performance was assessed using DSC, IoU, HSD95, and ASD. Better segmentation corresponds to larger DSC and IoU and smaller HSD95 and ASD. As in Chapter~3, the ``background'' and ``other'' categories were excluded because they do not identify specific brain structures and their already high measurements do not clearly quantify network performance.

\subsubsection{Single-Modal T1 Results}

Table~\ref{tab:acnn-ablation} presents ablation studies of the four proposed components on both datasets. Together with the parameter counts in Table~\ref{tab:parameter-comparison}, these results compare alternate with dense connectivity. On each dataset and for every metric, each successive architectural change on the left improves the corresponding measurements on the right. For a controlled comparison with Dolz et al.\cite{ref18}, the per-layer channel increment of DeepMedic was increased from 25 to 32, the smallest multiple of the basic ACNN channel count of 16 that is no smaller than 25. This modified DeepMedic was used as the baseline.

First, the OreoDown framework and its simplest instantiation were evaluated. DSC and IoU increased by approximately 1--2\% on both datasets, while HSD95 and ASD decreased rapidly, showing that OreoDown substantially improved contour accuracy. Second, the feature extractors of the instantiated downsampling process were implemented with dense or alternate connections to yield single-path DenseMedic and ACNN. Table~\ref{tab:parameter-comparison} shows that ACNN contains 2,976,064 parameters, 2.37 times fewer than DenseMedic's 7,046,464. Nevertheless, comparison of rows 3 and 4 in Table~\ref{tab:acnn-ablation} shows improved performance on all four IBSR metrics and three MRBrainS18 metrics. Better segmentation with fewer parameters supports the effectiveness of alternate connectivity. Relative to the second row, using alternate-connected FELs to instantiate OreoDown increased DSC by approximately 1.6\% and IoU by approximately 2.6\% on both datasets; HSD95 decreased by 3.5505 on IBSR and 0.3740 on MRBrainS18, while ASD decreased by 0.3598 and 0.1267, respectively.

Third, single-path ACNN was converted to multipath ACNN. Comparing rows 4 and 5 shows modest DSC and IoU improvements of 0.1--0.5\%, while HSD95 decreased by 0.8555 and 0.5620 and ASD by 0.1038 and 0.0147 on IBSR and MRBrainS18, respectively. The relatively large reductions in surface-distance metrics indicate that path division improves contour delineation on single-modal T1-MR images without changing the parameter count. Fourth, T1-MR images on different paths were spectrally enhanced. Comparing rows 5 and 6, DSC increased by 0.48\% and 0.09\%, IoU by 0.76\% and 0.16\%, HSD95 decreased by 0.7075 and 0.0507, and ASD decreased by 0.1025 and 0.0519 on IBSR and MRBrainS18, respectively. Both regional and contour accuracy therefore improved.

Overall, relative to the baseline, the final method increased DSC by 3.31\% on IBSR and 3.09\% on MRBrainS18, increased IoU by 5.12\% and 5.02\%, decreased HSD95 by 12.1781 and 5.2931, and decreased ASD by 1.1718 and 0.7791. Compared with the original DeepMedic, ACNN thus increased the spatial overlap between predicted and ground-truth structures and improved their contours. Lower standard deviations across images also indicate more robust segmentation.

\begin{table}[!htbp]
\centering
\caption{Ablation study of ACNN on (a) IBSR and (b) MRBrainS18. Values are mean $\pm$ standard deviation.}
\label{tab:acnn-ablation}
\scriptsize
\setlength{\tabcolsep}{1.8pt}
\resizebox{\textwidth}{!}{%
\begin{tabular}{@{}ccccccrrrr@{}}
\toprule
Baseline & New DS & DenseMedic & ACNN & Two-path & Spectral & DSC & IoU & HSD95 & ASD \\
\midrule
\multicolumn{10}{c}{\textit{(a) IBSR}} \\
$\checkmark$ & & & & & & $0.8639\pm0.0032$ & $0.7648\pm0.0045$ & $14.1656\pm2.5686$ & $2.0194\pm0.1950$ \\
$\checkmark$ & $\checkmark$ & & & & & $0.8753\pm0.0007$ & $0.7818\pm0.0012$ & $7.1010\pm1.6847$ & $1.4137\pm0.0389$ \\
$\checkmark$ & $\checkmark$ & $\checkmark$ & & & & $0.8870\pm0.0034$ & $0.8002\pm0.0053$ & $5.0719\pm2.2095$ & $1.2081\pm0.1394$ \\
$\checkmark$ & $\checkmark$ & & $\checkmark$ & & & $0.8912\pm0.0065$ & $0.8074\pm0.0103$ & $3.5505\pm1.4300$ & $1.0539\pm0.1124$ \\
$\checkmark$ & $\checkmark$ & & $\checkmark$ & $\checkmark$ & & $0.8922\pm0.0068$ & $0.8084\pm0.0110$ & $2.6950\pm0.6512$ & $0.9501\pm0.0323$ \\
$\checkmark$ & $\checkmark$ & & $\checkmark$ & $\checkmark$ & $\checkmark$ & $\mathbf{0.8970\pm0.0044}$ & $\mathbf{0.8160\pm0.0071}$ & $\mathbf{1.9875\pm0.1225}$ & $\mathbf{0.8476\pm0.0482}$ \\
\addlinespace
\multicolumn{10}{c}{\textit{(b) MRBrainS18}} \\
$\checkmark$ & & & & & & $0.8736\pm0.0036$ & $0.7770\pm0.0056$ & $6.5857\pm1.5270$ & $1.3016\pm0.1202$ \\
$\checkmark$ & $\checkmark$ & & & & & $0.8831\pm0.0021$ & $0.7928\pm0.0034$ & $2.2793\pm0.0977$ & $0.7158\pm0.0080$ \\
$\checkmark$ & $\checkmark$ & $\checkmark$ & & & & $0.8874\pm0.0019$ & $0.7996\pm0.0029$ & $1.3428\pm0.0211$ & $0.5902\pm0.0255$ \\
$\checkmark$ & $\checkmark$ & & $\checkmark$ & & & $0.9005\pm0.0043$ & $0.8208\pm0.0071$ & $1.9053\pm0.6339$ & $0.5891\pm0.0594$ \\
$\checkmark$ & $\checkmark$ & & $\checkmark$ & $\checkmark$ & & $0.9036\pm0.0026$ & $0.8256\pm0.0044$ & $1.3433\pm0.0749$ & $0.5744\pm0.0269$ \\
$\checkmark$ & $\checkmark$ & & $\checkmark$ & $\checkmark$ & $\checkmark$ & $\mathbf{0.9045\pm0.0011}$ & $\mathbf{0.8272\pm0.0018}$ & $\mathbf{1.2926\pm0.0233}$ & $\mathbf{0.5225\pm0.0100}$ \\
\bottomrule
\end{tabular}}
\end{table}

Table~\ref{tab:acnn-single-comparison} lists the four metrics obtained using leading FCN algorithms on IBSR and MRBrainS18. Each algorithm was reproduced according to its published architecture and parameters, with one exception: for HyperDense-Net, BN was replaced by GN with 16 groups to control the comparison with dual-path ACNN. To reduce or eliminate the influence of patch-size differences, DeepMedic and HyperDense-Net used input size 27 and output size 9, the same as ACNN. UNet and VoxResNet used $32\times32\times32$ patches, as close as practicable to ACNN's $27\times27\times27$ patches. Patch centers in their training and validation sets were identical to those used for ACNN.

\begin{table}[!htbp]
\centering
\caption{Algorithm comparison on single-modal T1 data from IBSR and MRBrainS18. Values are mean $\pm$ standard deviation.}
\label{tab:acnn-single-comparison}
\scriptsize
\setlength{\tabcolsep}{2.6pt}
\resizebox{\textwidth}{!}{%
\begin{tabular}{@{}lrrrrrrrr@{}}
\toprule
& \multicolumn{4}{c}{IBSR} & \multicolumn{4}{c}{MRBrainS18} \\
\cmidrule(lr){2-5}\cmidrule(l){6-9}
Method & DSC & IoU & HSD95 & ASD & DSC & IoU & HSD95 & ASD \\
\midrule
UNet\cite{ref12} & $0.8658\pm0.0072$ & $0.7685\pm0.0096$ & $6.0636\pm2.0927$ & $1.4571\pm0.1354$ & $0.8759\pm0.0105$ & $0.7820\pm0.0173$ & $2.0153\pm0.4307$ & $0.7561\pm0.0592$ \\
VoxResNet\cite{ref13} & $0.8847\pm0.0035$ & $0.7967\pm0.0052$ & $2.4277\pm0.5738$ & $0.9062\pm0.0707$ & $0.8743\pm0.0162$ & $0.7798\pm0.0260$ & $2.0476\pm0.4190$ & $0.7418\pm0.1512$ \\
DeepMedic\cite{ref18} & $0.8571\pm0.0027$ & $0.7544\pm0.0040$ & $21.5226\pm6.7981$ & $2.7501\pm0.4351$ & $0.9018\pm0.0008$ & $0.8229\pm0.0010$ & $1.8214\pm0.5022$ & $0.6145\pm0.0324$ \\
HyperDense-Net\cite{ref19} & $0.8841\pm0.0039$ & $0.7958\pm0.0060$ & $6.0435\pm1.0770$ & $1.3040\pm0.0578$ & $0.8975\pm0.0045$ & $0.8158\pm0.0073$ & $3.1746\pm1.3065$ & $0.7096\pm0.1425$ \\
ACNN & $\mathbf{0.8970\pm0.0044}$ & $\mathbf{0.8160\pm0.0071}$ & $\mathbf{1.9875\pm0.1225}$ & $\mathbf{0.8476\pm0.0482}$ & $\mathbf{0.9045\pm0.0011}$ & $\mathbf{0.8272\pm0.0018}$ & $\mathbf{1.2926\pm0.0233}$ & $\mathbf{0.5225\pm0.0100}$ \\
\bottomrule
\end{tabular}}
\end{table}

\begin{figure}[!htbp]
\centering
\includegraphics[width=\textwidth]{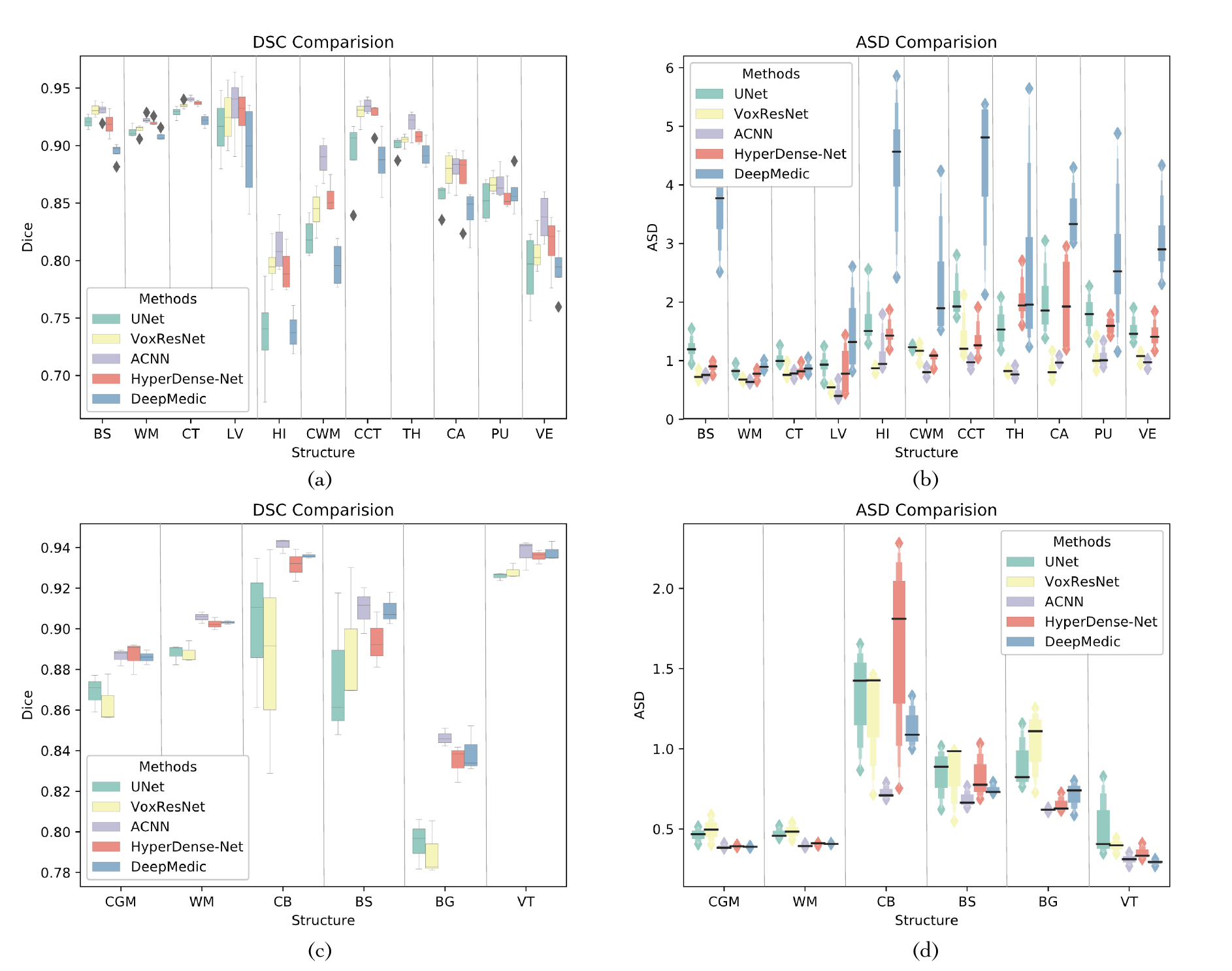}
\caption{Per-structure segmentation results of UNet, VoxResNet, ACNN, HyperDense-Net, and DeepMedic. IBSR abbreviations: brain stem (BS), cerebral white matter (WM), cerebral cortex (CT), lateral ventricle (LV), hippocampus (HI), cerebellar white matter (CWM), cerebellar cortex (CCT), thalamus (TH), caudate (CA), putamen (PU), and ventral diencephalon (VE). MRBrainS18 abbreviations: cortical gray matter (CGM), white matter (WM), cerebellum (CB), brain stem (BS), basal ganglia (BG), and ventricle (VT). (a) DSC on IBSR; (b) ASD on IBSR; (c) DSC on MRBrainS18; and (d) ASD on MRBrainS18.}
\label{fig:acnn-structures}
\end{figure}

ACNN achieved the best overall single-modal T1-MR performance in DSC, IoU, HSD95, and ASD on both datasets. On IBSR, relative to second-ranked VoxResNet, ACNN reached 89.70\% DSC, an increase of 1.23\%; 81.60\% IoU, an increase of 1.93\%; HSD95 of 1.9875, a reduction of 0.4402; and ASD of 0.8476, a reduction of 0.0586. On MRBrainS18, relative to second-ranked DeepMedic, ACNN reached 90.45\% DSC, an increase of 0.27\%; 82.72\% IoU, an increase of 0.43\%; HSD95 of 1.2926, a reduction of 0.5288; and ASD of 0.5225, a reduction of 0.092. Although the aggregate DSC gain was limited, the purple and red box plots in Figure~\ref{fig:acnn-structures}c show higher ACNN DSC than DeepMedic for every MRBrainS18 structure.

Figure~\ref{fig:acnn-structures} compares ACNN quantitatively with UNet, VoxResNet, DeepMedic, and HyperDense-Net for individual structures. Because of space constraints, the per-structure comparison uses only the principal DSC and ASD metrics. On the 11 IBSR structures in Figure~\ref{fig:acnn-structures}a and~b, ACNN achieved the leading DSC for every structure. For ASD, it ranked first in seven structures (WM, LV, CWM, CCT, TH, PU, and VE) and second in four (BS, CT, HI, and CA), while retaining the best overall average. On the six MRBrainS18 structures in Figure~\ref{fig:acnn-structures}c and~d, it achieved the leading DSC in five structures (WM, CB, BS, BG, and VT). Its DSC for cortical gray matter was slightly below HyperDense-Net, but the difference was very small and ACNN had lower variance and therefore greater robustness. ACNN achieved the leading ASD for every structure.

In summary, ACNN led every structure in IBSR DSC and MRBrainS18 DSC and ASD. In IBSR ASD, it led seven of 11 structures; its values were close to the leaders in the other four and its overall mean remained best. Compared with UNet, VoxResNet, HyperDense-Net, and DeepMedic, these results support the conclusion that ACNN produced more accurate regions and contours for each structure in single-modal T1-MR segmentation on both datasets.

\subsubsection{Dual-Modal T1+FLAIR and T1+T1-IR Results}

\begin{figure}[!htbp]
\centering
\includegraphics[width=\textwidth]{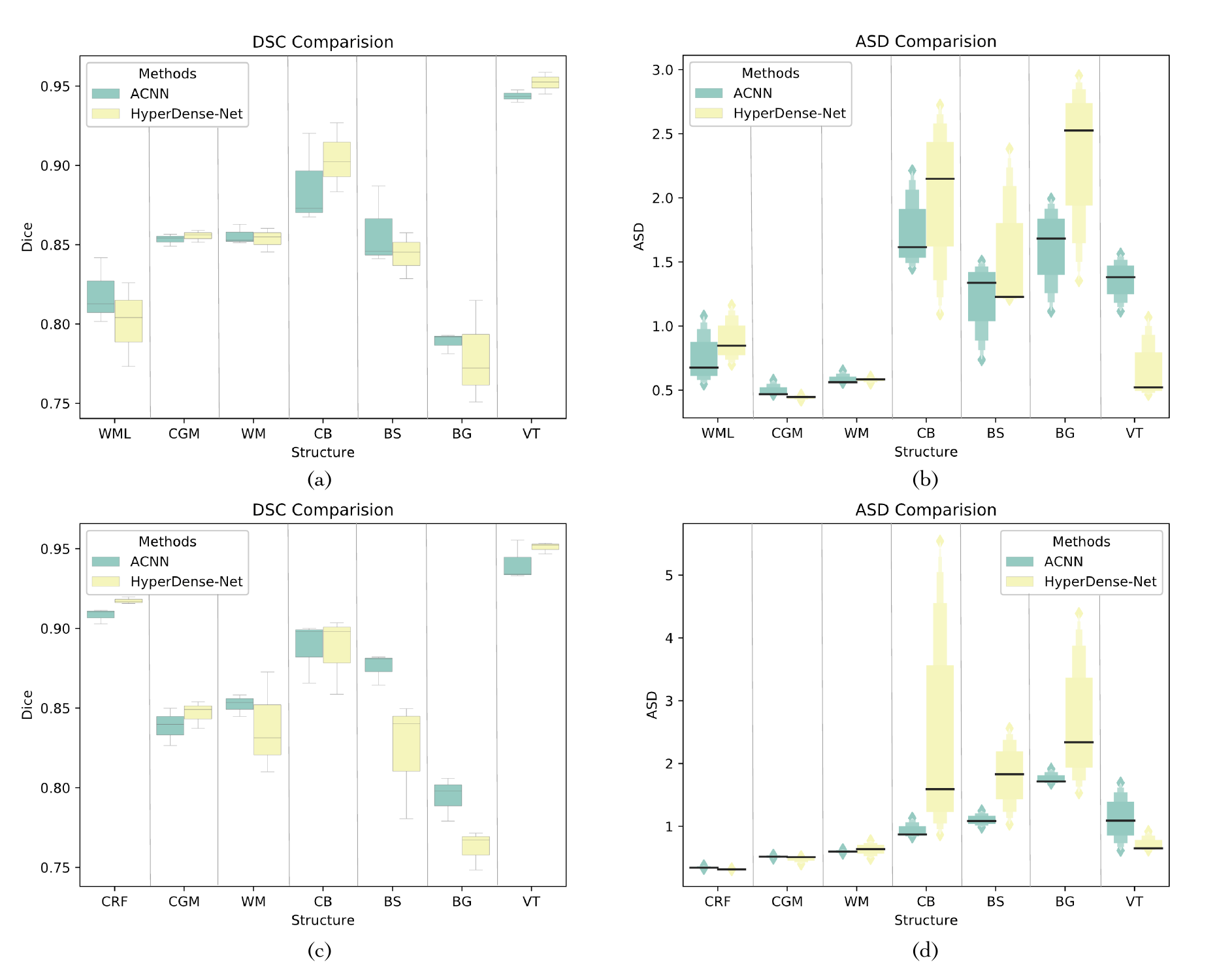}
\caption{Per-structure results of ACNN and HyperDense-Net on MRBrainS18. Abbreviations: white-matter lesion (WML), cerebrospinal fluid (CRF), cortical gray matter (CGM), white matter (WM), cerebellum (CB), brain stem (BS), basal ganglia (BG), and ventricle (VT). (a) DSC for T1+FLAIR; (b) ASD for T1+FLAIR; (c) DSC for T1+T1-IR; and (d) ASD for T1+T1-IR.}
\label{fig:acnn-dual-structures}
\end{figure}

ACNN's multimodal performance was evaluated using two dual-modal MRBrainS18 tasks. The T1+FLAIR task segmented white-matter lesions together with normal structures that are conspicuous in T1-MR, whereas the T1+T1-IR task segmented cerebrospinal fluid together with the T1-visible normal structures. Table~\ref{tab:t1-flair} and Figure~\ref{fig:acnn-dual-structures}a and~b present the T1+FLAIR results; Table~\ref{tab:t1-t1ir} and Figure~\ref{fig:acnn-dual-structures}c and~d present the T1+T1-IR results.

Dolz et al.\cite{ref19} had shown that HyperDense-Net outperformed UNet, VoxResNet, and DeepMedic on MRBrainS. The per-structure plots therefore compare only dual-path ACNN with HyperDense-Net, again using DSC and ASD. HyperDense-Net was reproduced using the published architecture and parameters, except that its BN was replaced by the same 16-group GN used in ACNN to reduce minibatch effects and improve control of the comparison. UNet, VoxResNet, and DeepMedic were reproduced according to their original descriptions. For these three single-path networks, multimodal MR patches were concatenated as input channels. Spectral enhancement was not used for ACNN in either dual-modal task, and the minibatch size remained 20.

On the T1+FLAIR task, dual-path ACNN achieved the best aggregate performance among all five algorithms. Relative to second-ranked HyperDense-Net, ACNN reached 85.78\% DSC, an increase of 0.22\%; 75.41\% IoU, an increase of 0.21\%; HSD95 of 5.2239, a reduction of 3.4247; and ASD of 1.1088, a reduction of 0.1026. For individual structures, ACNN led both DSC and ASD for white-matter lesions and basal ganglia; HyperDense-Net led both for cortical gray matter and ventricles. ACNN led ASD for white matter and cerebellum, whereas HyperDense-Net led DSC. ACNN led DSC for brain stem, whereas HyperDense-Net led ASD. The two networks therefore had different per-structure strengths, but ACNN showed slightly better overall performance.

\begin{table}[tbp]
\centering
\caption{Algorithm comparison on dual-modal T1+FLAIR data from MRBrainS18. Values are mean $\pm$ standard deviation.}
\label{tab:t1-flair}
\small
\resizebox{\textwidth}{!}{%
\begin{tabular}{@{}lrrrr@{}}
\toprule
Method & DSC & IoU & HSD95 & ASD \\
\midrule
UNet\cite{ref12} & $0.8169\pm0.0066$ & $0.6967\pm0.0096$ & $9.5785\pm2.1393$ & $1.3414\pm0.0791$ \\
VoxResNet\cite{ref13} & $0.8217\pm0.0091$ & $0.7036\pm0.0117$ & $10.9462\pm3.7014$ & $1.6001\pm0.3213$ \\
DeepMedic\cite{ref18} & $0.8261\pm0.0102$ & $0.7098\pm0.0140$ & $13.3767\pm3.6573$ & $1.8705\pm0.3985$ \\
HyperDense-Net\cite{ref19} & $0.8556\pm0.0086$ & $0.7520\pm0.0125$ & $8.6486\pm2.7315$ & $1.2114\pm0.1489$ \\
ACNN & $\mathbf{0.8578\pm0.0043}$ & $\mathbf{0.7541\pm0.0074}$ & $\mathbf{5.2239\pm1.9561}$ & $\mathbf{1.1088\pm0.1700}$ \\
\bottomrule
\end{tabular}}
\end{table}

On the T1+T1-IR task, dual-path ACNN again achieved the best aggregate performance among the five algorithms. Relative to second-ranked HyperDense-Net, ACNN reached 87.11\% DSC, an increase of 1.03\%; 77.46\% IoU, an increase of 1.4\%; HSD95 of 3.0065, a reduction of 5.0837; and ASD of 0.9170, a reduction of 0.4228. For individual structures, HyperDense-Net led both DSC and ASD for cerebrospinal fluid, cortical gray matter, and ventricles. ACNN led both metrics for white matter, cerebellum, brain stem, and basal ganglia. ACNN therefore led four of the seven structures and had better overall performance.

\begin{table}[tbp]
\centering
\caption{Algorithm comparison on dual-modal T1+T1-IR data from MRBrainS18. Values are mean $\pm$ standard deviation.}
\label{tab:t1-t1ir}
\small
\resizebox{\textwidth}{!}{%
\begin{tabular}{@{}lrrrr@{}}
\toprule
Method & DSC & IoU & HSD95 & ASD \\
\midrule
UNet\cite{ref12} & $0.3505\pm0.0247$ & $0.2455\pm0.0264$ & $17.4414\pm4.6833$ & $4.8032\pm1.0171$ \\
VoxResNet\cite{ref13} & $0.4093\pm0.1076$ & $0.2911\pm0.0896$ & $13.3716\pm5.2617$ & $3.4607\pm1.2250$ \\
DeepMedic\cite{ref18} & $0.5828\pm0.0901$ & $0.4290\pm0.0878$ & $18.5431\pm3.4524$ & $3.7601\pm1.3313$ \\
HyperDense-Net\cite{ref19} & $0.8608\pm0.0130$ & $0.7606\pm0.0192$ & $8.0902\pm1.5278$ & $1.3398\pm0.2357$ \\
ACNN & $\mathbf{0.8711\pm0.0075}$ & $\mathbf{0.7746\pm0.0117}$ & $\mathbf{3.0065\pm0.8874}$ & $\mathbf{0.9170\pm0.0872}$ \\
\bottomrule
\end{tabular}}
\end{table}

Across the two dual-modal tasks, ACNN and HyperDense-Net each had per-structure advantages, with ACNN achieving a slight overall lead and better mean DSC, IoU, HSD95, and ASD. ACNN contains 2,976,064 parameters, as shown in Table~\ref{tab:parameter-comparison}, whereas HyperDense-Net contains 10,349,450\cite{ref19}. ACNN therefore reduced the parameter count by a factor of 3.48 while outperforming the densely connected HyperDense-Net, further demonstrating the effectiveness of alternate connectivity in dual-modal segmentation. The results support the conclusion that ACNN achieved advanced performance on both T1+FLAIR and T1+T1-IR segmentation on MRBrainS18.

\subsection{Visual Assessment of Subcortical Segmentation}

To provide a more direct impression of the quantitative results and qualitatively assess the networks, this subsection presents selected segmentations. Because ACNN belongs to the patch-wise downsampling-FCN family, it is compared visually with DeepMedic and HyperDense-Net. Four tasks are shown: IBSR T1-MR, MRBrainS18 T1-MR, MRBrainS18 T1+FLAIR, and MRBrainS18 T1+T1-IR.

\subsubsection{Single-Modal T1 Segmentation}

\begin{figure}[!htbp]
\centering
\includegraphics[width=\textwidth]{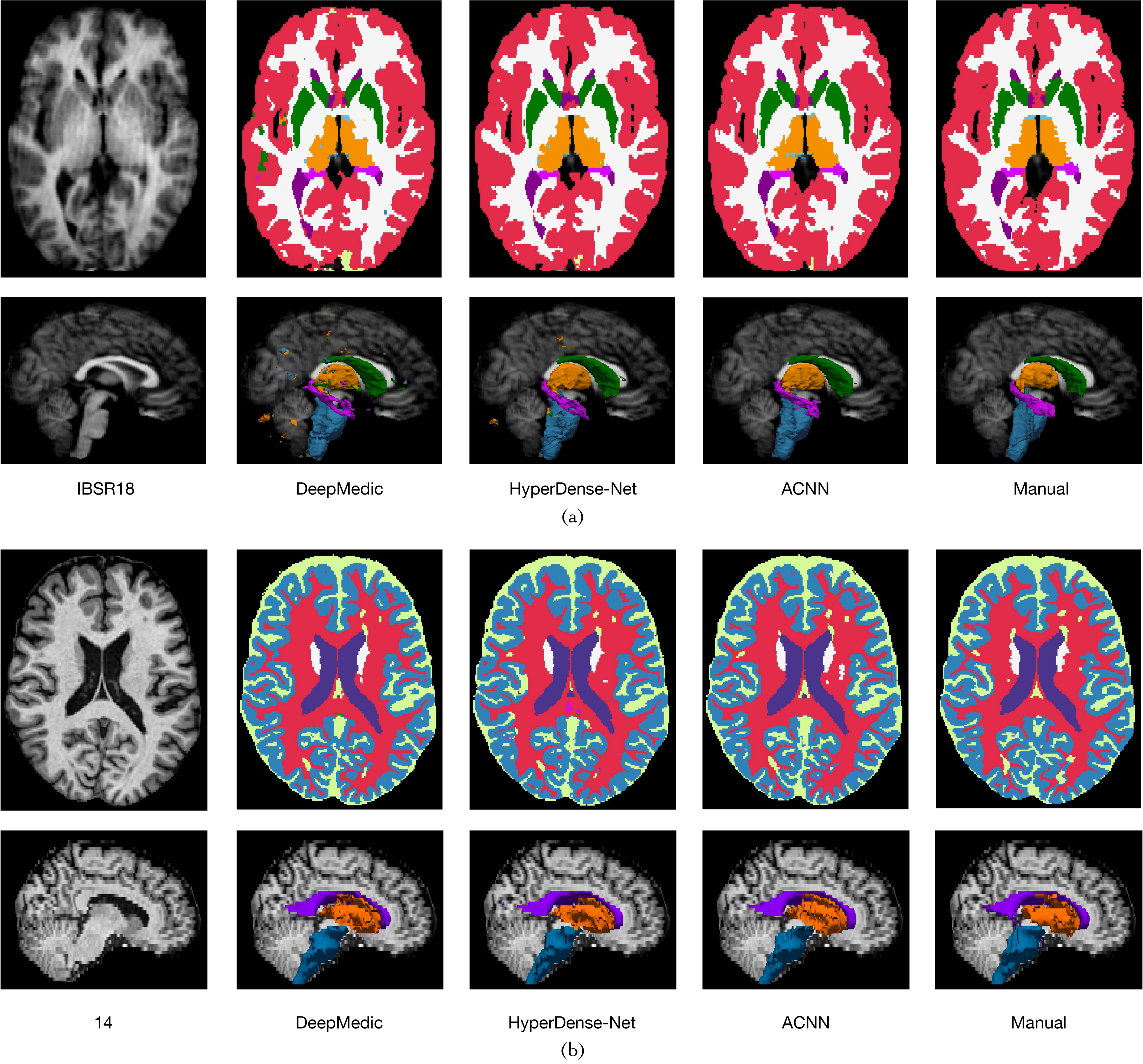}
\caption{Visual assessment of single-modal segmentation: (a) IBSR T1-MR and (b) MRBrainS18 T1-MR.}
\label{fig:acnn-single-visual}
\end{figure}

Figure~\ref{fig:acnn-single-visual} shows selected single-modal results of DeepMedic, HyperDense-Net, and ACNN. In Figure~\ref{fig:acnn-single-visual}a, the first row shows planar views of selected IBSR structures in the axial plane, and the second row shows three-dimensional sagittal views of the caudate, thalamus, hippocampus, and brain stem. The ACNN planar segmentation most closely resembles the ground truth in the morphology of each structure. In the three-dimensional views, DeepMedic produces false-positive fragments in the yellow thalamus, green caudate, and blue brain stem, together with a clear defect in the center of the magenta hippocampus. HyperDense-Net improves these findings but retains a few fragments in the thalamus and caudate and a cavity in the hippocampus. ACNN eliminates the false-positive fragments in the three structures and the hippocampal defect, thereby extending the DeepMedic family under the same input--output relationship.

In Figure~\ref{fig:acnn-single-visual}b, the first row shows an axial plane from MRBrainS18 case 14, and the second row shows sagittal three-dimensional forms of the basal ganglia, ventricle, and brain stem. In the planar view, ACNN delineates the cerebrospinal fluid, gray matter, and white matter more accurately and produces fewer misclassifications. Its three-dimensional forms also more closely resemble the ground truth. DeepMedic performs better than the larger HyperDense-Net in this MRBrainS18 plane, possibly because fewer parameters facilitate rapid convergence on a small dataset. Overall, ACNN produces more accurate planar and three-dimensional results in the single-modal tasks.

\subsubsection{Dual-Modal T1+FLAIR and T1+T1-IR Segmentation}

\begin{figure}[!htbp]
\centering
\includegraphics[width=\textwidth]{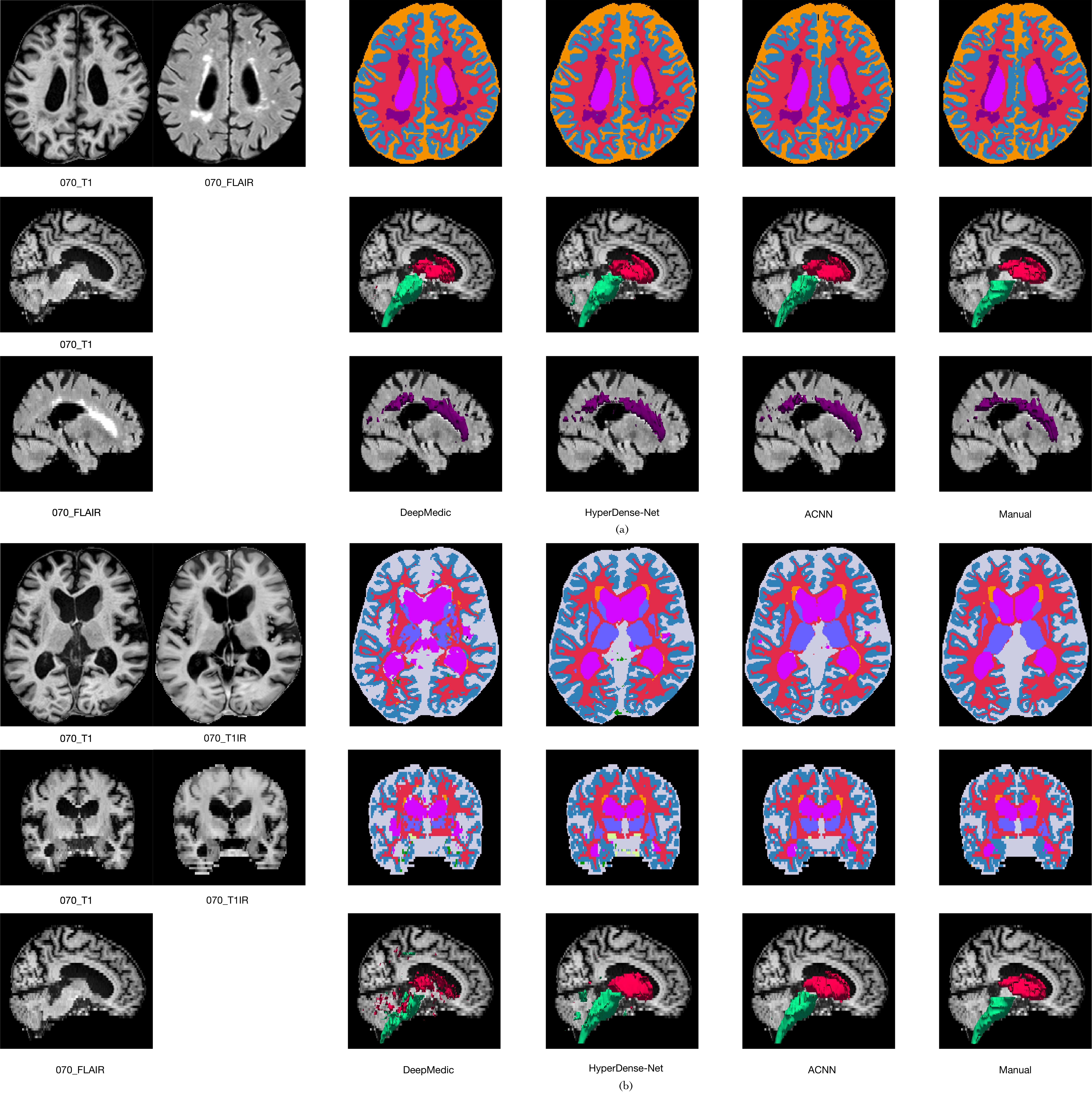}
\caption{Visual assessment of dual-modal segmentation: (a) MRBrainS18 T1+FLAIR and (b) MRBrainS18 T1+T1-IR.}
\label{fig:acnn-dual-visual}
\end{figure}

Figure~\ref{fig:acnn-dual-visual} shows selected dual-modal segmentations of MRBrainS18 case 070 by DeepMedic, HyperDense-Net, and ACNN. For T1+FLAIR in Figure~\ref{fig:acnn-dual-visual}a, the first row shows an axial plane, the second row shows sagittal three-dimensional forms of the T1-based basal ganglia and brain stem, and the third row shows the three-dimensional FLAIR-based white-matter-lesion region. In the first row, all three networks produce similar segmentations of the magenta T1-based ventricle, red white matter, and cyan gray matter; yellow denotes the ``other'' class. For the purple FLAIR-based white-matter lesion, the ACNN result remains discontinuous on the left, as do the other two results, but most closely matches the ground-truth morphology on the right. In the second row, DeepMedic produces false-positive fragments in the red basal ganglia and HyperDense-Net in the cyan brain stem, whereas ACNN produces neither. In the third row, ACNN produces the smallest lesion fragments and the closest correspondence between the left and right components.

For T1+T1-IR in Figure~\ref{fig:acnn-dual-visual}b, the first row shows one axial plane of case 070, the second row one coronal plane, and the third row three-dimensional forms of the T1-based red basal ganglia and cyan brain stem. T1 and T1-IR are used to delineate the outer boundary of cerebrospinal fluid in MRBrainS18. A three-dimensional rendering of this structure would enclose the entire brain, so cerebrospinal-fluid segmentation is shown only in axial and coronal planes. Because T1 and T1-IR carry similar semantic information, supplying their simple two-channel concatenation to a single-path network may introduce redundancy and impair segmentation. This helps explain the low measurements of DeepMedic, UNet, and VoxResNet and the poor visual DeepMedic result in Figure~\ref{fig:acnn-dual-visual}b.

Among the better-performing HyperDense-Net and ACNN, the first two rows show that ACNN more closely matches the ground truth for T1-based tissues and produces a cleaner cerebrospinal-fluid region apart from limited encroachment by other tissues. In the third row, ACNN's basal-ganglia and brain-stem results also avoid the fragments present in HyperDense-Net. Thus, ACNN produces more accurate planar and three-dimensional segmentations in both dual-modal tasks. Considered together with the single-modal results, the proposed ACNN achieves more accurate and robust segmentation across all evaluated tasks with relatively few parameters.

\section{Chapter Summary}

To improve regional and contour accuracy in T1-MR subcortical segmentation while also achieving advanced multimodal performance, this chapter proposed ACNN within the abstract OreoDown architecture. Alternate-connected FELs and downsamplers instantiate OreoDown as ACNN, improving segmentation while reducing the parameter count by several fold. The single path is then divided into multiple paths without changing the parameter count, improving T1-MR segmentation while supporting advanced multimodal segmentation with a lightweight model. Finally, prior information is introduced into multipath ACNN for single-modal segmentation, producing a further improvement on T1-MR images.

Experiments on IBSR and MRBrainS18 showed that every modification improved regional and contour accuracy on single-modal T1-MR images. Compared with previous methods, ACNN performed strongly on all four conventional metrics in both single- and dual-modal segmentation. Its subcortical regions were more accurate and its contours more regular, as also shown by the visual assessments. In clinical applications, precise subcortical segmentation may support accurate measurement of key diagnostic indicators and rapid computer-assisted treatment.

\chapter{Conclusions and Future Work}

\section{Summary of the Thesis}

This thesis first conducted a detailed review and analysis of semantic segmentation of subcortical structures in MR images and used that analysis to define the research scope and objectives. A technical workflow was designed for the selected datasets, deep learning-based segmentation algorithms were developed, and their performance was quantified using standard segmentation metrics. The planned research was substantially completed, meeting the objectives of a large receptive field, high performance with few parameters, and a unified approach to single- and multimodal segmentation. The principal work is summarized as follows.

\begin{enumerate}
\item \textbf{Inspired by the calculation of receptive fields, the growth rate of the receptive field was increased at an early stage and network depth was preserved with a sandwich-like arrangement, producing an effective receptive-field expansion.}

Under the input--output relationship of a downsampling FCN, the stride of an early convolution was increased to accelerate receptive-field growth. Size-preserving convolutional layers were then inserted in a sandwich-like arrangement to restore depth, converting the higher growth rate into an effective receptive-field increase. This led to OreoDown, an abstract, sandwich-shaped downsampling framework with a large receptive field.

\item \textbf{Two inherited multiscale information-fusion schemes, dense connectivity and alternate connectivity, were designed for single-modal tasks.}

On the basis of the instantiated downsampling layers, the feature-extraction component of OreoDown was redesigned. Modules composed of multiple densely connected feature-extraction layers instantiate OreoDown as DenseMedic. Alternate connections between layers were then proposed, and multiple alternate-connected units instantiate OreoDown as ACNN.

\item \textbf{The alternate connected neural network, which achieved advanced single-modal performance, was extended to achieve advanced performance in multimodal segmentation.}

The single-modal performance of both networks was first extended and strengthened. Hybrid dilated convolution was introduced into DenseMedic to provide refined perception over a large receptive field. ACNN was divided into multiple paths without changing the parameter count, and effective prior information was introduced into the multipath network under single-modal input to vary its input distributions. These changes improved single-modal segmentation. Multipath ACNN also achieved advanced multimodal performance, thereby unifying single- and multimodal segmentation under a high-performing architecture.
\end{enumerate}

Supported by the experimental results, the principal conclusions are as follows.

\begin{enumerate}
\item \textbf{Receptive-field expansion.} Under the input--output relationship of a downsampling FCN, increasing the stride of an early convolution accelerates receptive-field growth. Restoring depth with sandwich-like size-preserving convolutions can effectively improve segmentation performance.

\item \textbf{Reduction in parameter count.} Dense and alternate connectivity provide two multiscale information-fusion approaches for single-modal tasks. At the evaluated depth, ACNN uses 2.37 times fewer parameters than DenseMedic while improving single-modal segmentation.

\item \textbf{Unified single- and multimodal segmentation.} Cross-path alternate connectivity, which performs strongly in single-modal segmentation, also performs strongly in multimodal segmentation. Introducing prior information into the single-modal inputs of multipath ACNN effectively addresses the tendency of different paths to receive identical intensity distributions. Because multipath construction adds no parameters, redundant information from the same input on multiple paths is not a major problem; nevertheless, differentiating the intensity distributions and adding semantic information further improves single-modal performance.
\end{enumerate}

Having completed the research and experiments and established these conclusions, the principal innovations of the thesis are as follows.

\begin{enumerate}
\item \textbf{Receptive field.} Beginning from the calculation of a convolutional network's receptive field, the thesis derives a means of expanding it and demonstrates a corresponding improvement in segmentation. The network architecture is therefore designed and validated from a mathematical basis, making the receptive-field expansion interpretable and partially reducing the black-box character of the neural network.

\item \textbf{Neural-network architecture.} Building on receptive-field expansion, the thesis primarily improves segmentation by designing the network architecture. The proposed alternate connectivity achieves high-performance segmentation in both single- and multimodal tasks with relatively few parameters and can provide a reliable, high-performing baseline for higher-level mechanisms such as attention.

\item \textbf{Prior information.} Prior information is designed to vary the distribution of a single image across paths. Introducing additional semantic information from image space and texture further improves single-modal segmentation. A single-modal image can then be treated as a distinctive ``multi''-modal task, allowing the same network to unify single- and multimodal segmentation.
\end{enumerate}

\section{Future Work}

Based on further consideration and the most recent literature available at the time of the study, future work may continue at both the technical and clinical levels.

\textbf{Technical directions.} Further methods of introducing more effective prior information can be explored for T1-only segmentation with multipath ACNN, together with additional network instantiations within the large-receptive-field OreoDown framework. Larger datasets can be collected and annotated to derive general pretrained parameters for OreoDown instantiations and to investigate low-cost transfer learning from source to target domains. Meta-learning and metric learning may support direct partitioning of feature space and transfer of information between domains. On top of FCN architecture design represented by OreoDown instantiations, attention mechanisms and loss-function designs can be integrated in deep layers to improve segmentation further. Distance transforms may also improve contour-oriented metrics such as HSD95 and ASD.

\textbf{Clinical directions.} Accurate segmentation of subcortical structures can support diagnosis and treatment of several brain diseases. In autism spectrum disorder, morphological and volumetric changes in the caudate and putamen are closely related to diagnosis. Precise segmentation by DenseMedic and ACNN may improve measurement of these structures and support assessment of possible restricted and repetitive behaviors. During radiotherapy for brain cancer, clinicians manually delineate target volumes and organs at risk from brain images to concentrate radiation in the target while minimizing unnecessary exposure of surrounding normal tissues and organs. This voxel-by-voxel delineation is time-consuming. Accurate segmentation of normal brain regions by DenseMedic and ACNN may help optimize the physical dose distribution and its spatial placement, support automatic delineation of radiotherapy targets, reduce treatment costs, and save valuable treatment time. The segmentation improvements produced by the two proposed networks therefore also have potential clinical significance.

\backmatter
\renewcommand{\bibname}{References}

\chapter*{Acknowledgments}
\addcontentsline{toc}{chapter}{Acknowledgments}

This thesis was completed under the kind care and meticulous guidance of my respected supervisor, Associate Researcher Weiwei Zhang. His serious scientific attitude, rigorous scholarship, and commitment to excellence have profoundly influenced and inspired me. From the selection of the research topic to the completion of the thesis, Professor Zhang consistently provided careful guidance and unstinting support. I offer him my sincere gratitude and highest respect.

I would also like to thank Professor Haiyan Xu, Researcher Yi Peng, Researcher Jian Liu, and Linwen Liu of the School of Basic Medicine, Peking Union Medical College, as well as Yixuan Cao of the Institute of Computing Technology, Chinese Academy of Sciences. Their help and support enabled me to overcome successive difficulties and uncertainties and to complete this thesis successfully.

As the thesis draws to a close, it is difficult to remain unmoved when recalling how many respected teachers, classmates, and friends offered their quiet assistance from the beginning of the project through its completion. Please accept my sincere thanks. Thank you all.

\end{document}